\documentclass{emulateapj}
\usepackage{graphicx}
\usepackage{longtable}
\usepackage{natbib}
\usepackage{apjfonts}
\usepackage{subfig}
\newcommand{\mj}{$M_{\mathrm{J}}$}
\newcommand{\rj}{$R_{\mathrm{J}}$}
\newcommand{\me}{$M_{\oplus}$}

\newcommand{\qp}{$Q_{\mathrm{p}}'$}
\newcommand{\qs}{$Q_{\mathrm{s}}'$}

\newcommand{\cp}{\citep}
\newcommand{\ct}{\citet}

\def\apjs{ApJS}
\def\apjl{ApJ}
\def\aap{A\&A}

\begin{document}
\title{Inflating and Deflating Hot Jupiters: \\
  Coupled Tidal and Thermal Evolution of Known Transiting Planets}
\author{N. Miller\altaffilmark{1}, J. J. ~Fortney\altaffilmark{1}, and B. Jackson\altaffilmark{2}}
\altaffiltext{1}{Department of Astronomy and Astrophysics, University of California, Santa Cruz 
  (neil@astro.ucsc.edu, jfortney@ucolick.org)}
\altaffiltext{2}{Lunar and Planetary Laboratory, University of Arizona (\mbox{bjackson}@lpl.arizona.edu)}

\begin{abstract}
{We examine the radius evolution of close-in giant planets with a planet evolution model 
that couples the orbital-tidal and thermal evolution.  
For 45 transiting systems, we compute a large grid of cooling/contraction paths forward in time, 
starting from a large phase space of initial semi-major axes and eccentricities.  
Given observational constraints at the current time for a given planet 
(semi-major axis, eccentricity, and system age) we find possible evolutionary paths that match these constraints, 
and compare the calculated radii to observations.  We find that tidal evolution has two effects.  
First, planets start their evolution at larger semi-major axis, 
allowing them to contract more efficiently at earlier times.  
Second, tidal heating can significantly inflate the radius when the
orbit is being circularized, but this effect on the radius is short-lived thereafter.  
Often circularization of the orbit is proceeded by a long period while the semi-major axis slowly decreases.
Some systems with previously unexplained large radii that we can reproduce with our coupled model are 
HAT-P-7, HAT-P-9, WASP-10, and XO-4.  This increases the number of planets for which we can match the radius from 24 (of 45) 
to as many as 35 for our standard case, but for some of these systems we are required to be viewing them at a 
special time around the era of current radius inflation.  
This is a concern for the viability of tidal inflation as a general mechanism to explain most inflated radii.  
Also, large initial eccentricities would have to be common.
We also investigate the evolution of models that have a floor on the eccentricity, as may be due to a perturber. 
In this scenario we match the extremely large radius of WASP-12b.  
This work may cast some doubt on our ability to accurately determine the interior heavy element enrichment of normal, 
non-inflated close-in planets, 
because of our dearth of knowledge about these planet's previous orbital-tidal histories.
Finally, we find that the end state of most close-in planetary systems is disruption of the planet as it moves ever closer to its parent star.}
\end{abstract}

\section{Introduction}
The precise mass and radius measurements for transiting exoplanets provide information about the planets' 
interior structure and composition, which are often apparently unlike that of Jupiter and
Saturn.  Indeed, it is the incredible diversity of measured radii of transiting planets that has been most
surprising.  In the solar system, Jupiter and Saturn differ in mass by a factor of three while their radii
differ by only 18\%.  However, amongst exoplanets, planets with the same mass can differ in radius by a
factor of two.  A hope amongst planetary astrophysicists was that the measurement of the mass and radius, when
compared to models, would cleanly yield information on planetary interior composition.  Although there are clearly
examples where this has been done successfully, including heavy element rich planets such as HD 149026b
\citep{Sato05,Fortney06} and GJ 436b \citep{Gillon07}, in general modelers have been foiled by planets with very large
radii, larger than can be accommodated by ``standard'' cooling/contraction models.

Considerable work has been done in the past several years to understand the large radii of some planets, 
as well as the radius distribution of the planets as a whole.  Explanations for the ``anomalously'' large 
planets have fallen into three categories:  those that are a current or recent additional internal energy source, 
which has stalled the interior
cooling and contraction \citep{Bodenheimer01,Guillot02,Bodenheimer03,Gu03,Winn05,LiuBurrowsIbgui08,Jackson08,IbguiBurrows2009}, 
those that instead merely delay the contraction by slowing the transport of interior
energy \citep{Burrows07,Chabrier07c}, and those that invoke various evaporation mechanisms
\citep{Baraffe04,Hansen07}.  These are briefly reviewed in \citet{Fortney08b}.

Tidal heating as an explanation for these large-radius planets was suggested by \citet{Bodenheimer01} for HD 209458b and
has been revisited frequently by other authors \cp[e.g.][]{Bodenheimer03,Winn05,LiuBurrowsIbgui08,Gu03,Gu04,Jackson08a,Jackson08,IbguiBurrows2009}.  
We note that the mechanism of heating by obliquity tides \citep{Winn05} has been cast in considerable double by several authors
\citep{Levrard07, FabryckyTremaine2007, Peale2008}.

At this time, tidal heating by orbit circularization is generally believed to be the most important type.
The largest uncertainties in the standard tidal theory is the ``tidal $Q$'' value, 
 a standard parameterization of the rate of tidal effects.
In this work, we use the standard notation for the planet tidal $Q$ value as \qp\ and the stellar tidal $Q$ value as \qs\ .
Jupiter's \qp\ value has been constrained to be between  $10^5$ and $10^{6}$ \citep{Goldreich66}.
For tidal heating by circularization to take place, 
the planet must either initially have an eccentric orbit or the system must be driving the eccentricity of the planet at recent times.  

The former scenario would have the following qualitative stages.   
The planet is left with an eccentric orbit through planet-planet interactions \citep{RasioFord96,
ChatterjeeFordMatsumuraRasio2008,FordRasio2008}.
Tides on the star gradually reduce the semi-major axis. These tidal effects accelerate as the semi-major axis decreases.  
Tides on the planet become more important and the planet's orbit circularizes; at the same time depositing
orbital energy into the planet's interior.  
Scattering/tidal evolution models of this sort were recently computed by \ct{Nagasawa08}.
At this point, the system might be observed to have a fairly circular orbit and a larger-than-expected radius.
\citet{IbguiBurrows2009} use a coupled tidal-thermal evolution model, quite similar to the one we present here, 
to show that this scenario 
might be possible for the HD 209458 system, and by extension, many hot Jupiter planets.  
Such a model is necessary to self-consistently explain
a planet's radius in this picture.  One potential issue with this scenario is that it can require
large-radius planets to be observed at a ``special time'' since after the orbit is circularized, the planet
may rapidly contract.

Alternatively, some planets might be found in an equilibrium state where their eccentricity is being forced by a third body
while at the same time tides on the planet are damping the eccentricity \citep{Mardling2007}.  
This is an attractive explanation because the planet might be found in an inflated state for a long period of time.
Previously,
\citet{Bodenheimer01} calculated the tidal power required to maintain the radius for HD 209458 b, Ups And b, and Tau Boo b, as a function
of the assumed core size, in a stationary orbit.
Recently, thermal evolution calculations with constant heating have been performed for TrES-4, XO-3b and HAT-P-1b by 
\citet{LiuBurrowsIbgui08}, who placed constraints on $\bar{e}^2/Q_p'$ - where $\bar{e}$ is the recent time-averaged eccentricity of the orbit.   
These calculations are useful for estimating the required recent tidal heating.
In some cases, where the eccentricity is non-zero and a perturber is necessary to invoke, then
this constant heating picture might accurately describe the recent thermal history of the planet.
In many cases the eccentricity is observed to be close to zero, which either implies that
a) the planet's eccentricity is at a non-zero equilibrium, but the planet's \qp\ value is much smaller
than inferred from Jupiter or b) the planet's orbit is circularized and this calculation does not apply.

Clearly it is important to accurately measure the eccentricity of inflated systems to determine
if either scenario is plausible.  For many transiting systems, the eccentricity has been only 
weakly constrained with several radial velocity points and it is very difficult to 
distinguish a small eccentricity from one that is truly zero \cp{Laughlin05b}.  
For systems with an observed secondary eclipse, stronger upper limits on eccentricity 
can be found based on the timing of the eclipse \cp{Deming05b,Charb05,Knutson09}.  Note that
secondary eclipse timing only constrains $e \cos{\Omega}$ so it is possible that some of these
systems have much larger eccentricity, but it is unlikely.

The above possibilities are also consistent with the popular planet formation and migration theories.
These planets form while the protoplanetary disk is still present at much larger 
orbital distances and migrate early in their life to small orbital distances \cp[e.g.][]{Lin96}.  
After this initial phase, tidal evolution between the star and the planet occurs on Gyr time scales.
The migration mechanism is important because it determines the initial orbital parameters for 
tidal evolution.  There are multiple postulated migration mechanisms.
\begin{enumerate}
\item
  Planet-disk interaction: Gravitational interactions between the planet and protoplanetary disk
  can exert torque on the planet \cp{Ward97a,Ward97b}.  
  These mechanisms tend to circularize the planet's orbit very early on and decrease the semi-major axis.  
  The disk migration time scales is significantly
  shorter than the lifetime of the disk, as described in \citet{Papaloizou07} and references therein.
\item
  Planet-planet interaction: Gravitational interactions with other nearby planets
  can transfer orbital energy and angular momentum between the two bodies.  This can result
  in quickly decreasing or increasing the orbital distance of one of the planets as
  well as producing non-zero initial eccentricity orbits.  
  Using N-body simulations, \citep{RasioFord96, Weidenschilling96, ChatterjeeFordMatsumuraRasio2008, FordRasio2008} 
  have shown that this effect
  can be important and can result in the inner bodies having initial eccentricity as large as 0.8, before tidal damping ensues.
  Other authors have investigated migration with coupled secular driving and tidal friction, which can operate on similar timescales
  \citep{WuMurray03,Faber05,FordRasio2006,FabryckyTremaine2007}.
  There are also a handful of transiting planets that have non-zero eccentricity today, which can be
  explained by planet-planet interactions.  It is also suggestive that many of these
  eccentric planets are more massive and have longer circularization time scales.
  Since the circularization time is longer for massive planets, this observation is consistent
  with the idea that planets of all masses can have large initial eccentricity, but that the
  lower mass planets have circularized while the massive planets may still be circularizing.
\end{enumerate}
We expect that both of these mechanisms do happen to some extent.  
Therefore, we assume that a wide range of initial orbital parameters are possible, 
and we following the orbital and structural evolution of planets from a wide range of possible initial eccentricities, 
as described below.  
In the absence of a theory to predict likely initial eccentricities for a given planetary system, 
we seek to understand the physics of the evolution from a variety of initial states.

Most of the detected transiting planets currently have small eccentricities consistent with zero.
These can be explained by either migration mechanism.  If the planet migrated through
planet-disk interactions, then it would have zero eccentricity when tidal evolution began.  If
the planet migrated through planet-planet interactions, then the orbit may have circularized
due to tides on the planet.  

\section{Model: Introduction}
In this work we would like to test the possibility that tidal heating by orbit circularization 
can explain the transit radius observations for each particular system.  A necessary condition
for this model is that a self-consistent evolution history can be found that agrees
with all of the observed system parameters.
To check this condition, we forward-evolve a coupled tidal-thermal evolution model over a large grid of initial semi-major axis
and eccentricity for each system.  We perform this test for $Q_{\mathrm{s}}'=10^5$, $Q_{\mathrm{p}}'=10^5$ and 
$Q_{\mathrm{s}}'=10^5$, $Q_{\mathrm{p}}'=10^{6.5}$.  Also, for each system with non-zero current eccentricity, we emulate
an eccentricity driving source by performing runs with an eccentricity floor equal to the observed value.  Later we also explore
some higher \qs\ cases.

%% SOME DETAILS ABOUT VARIOUS STAGES && EFFECTS
To properly understand the planet's thermal evolution, it is necessary to couple the planet thermal
evolution to the orbital-tidal evolution.  
Generally planets with initial semi-major axis of 0.1 AU or less will spiral into the star in Gyr timescales
\citep{Jackson08a}.  
This has
a large impact on the incident flux on the planet, and therefore the loss of intrinsic luminosity of the planet.  
For some systems, this more efficient cooling at early times makes it possible to achieve smaller
radii at the present.
As the planet moves closer to the star, the tidal effects accelerate.
If the orbit is eccentric, then at some point the planet's orbit undergoes a period of circularization.
At this time a significant amount of orbital energy is deposited into the planet, which 
increases its radius.  The question of this work is whether, at this stage, the system's observables ($a$, $e$, age, $R$)
can simultaneously be achieved in the model.
After this stage, the planet may lose mass by Roche lobe overflow \citep{Gu03}, which can temporarily
prevent the planet from falling into the star.  However, the planet's destiny is to fall into the star 
\citep{Levrard09,Jackson09}.
These final stages of the planet's life, including the mass loss stage, are not modeled in this work.

We typically find
that tides on the star are the dominant source of semi-major axis
evolution \cp{Jackson08a,Jackson08}.  When the eccentricity is large and damping, the tides on the planet can be the dominant
semi-major axis damping source \citep{Jackson08a,IbguiBurrows2009}.  
After surveying our suite of systems, 
we find that tidal heating can usually provide sufficient
energy to inflate planetary radii as large as observed, but we do not 
always find an evolutionary history where the radius, semi-major axis, eccentricity and age
all simultaneously fall within the observed error bars.  
Regardless, we find that tidal processes are an important
aspect of planet evolution, particularly for hot Jupiter systems.

\section{Model: Implementation}
The \citet{Fortney07a} giant planet thermal evolution model has been coupled
to the \citet{Jackson08} tidal evolution model.  Therefore, the 
semi-major axis, eccentricity, and radius of the planet all evolve simultaneously.  
The tidal power is assumed to be deposited uniformly into the envelope of the planet.  
The planet structure model is assumed to be composed of four parts:
\begin{enumerate}
\item
  a  50\% rock/ 50\% ice core (by mass) with the ANEOS equations of state \citep{ANEOS}.  
  The core does not participate in the thermal evolution of the planet, as in \ct{Fortney07a}.
\item
  a H/He envelope with $Y=0.27$, which uses the equation of state of \citet{SCVH}.  The envelope
  is assumed to be fully convective and thus has constant specific entropy throughout.  
  At each time step the envelope is assumed
  to be in hydrostatic equilibrium.
\item
  a series of radiative-convective, equilibrium chemistry, non-grey atmosphere models described
  in more detail in \citet{Fortney07a} and \ct{Fortney08a}.  
  These grids are computed for the incident fluxes at 0.02, 0.045, 0.1, and 1 AU from the Sun. 
  This correctly determines the atmospheric structure and luminosity of the planet as a function of the planet's surface gravity,
  incident flux from the host star, and interior specific entropy.  In cases where the planet migrates to a semi-major axis 
  with more incident flux than the innermost grid, then the boundary condition at the innermost grid is used.
\item
  an extension of the atmosphere to a radius where the slant optical depth in a wide optical band 
  (the \emph{Kepler} bandpass) reaches unity.  
  Therefore, all plotted radii are at the ``transit radius,'' as discussed by several authors 
  \cp{Hubbard01,Baraffe03,Burrows03}.  
  The slant optical depth as a function of pressure is computed with the code described in \ct{Hubbard01} and \ct{Fortney03}.
  We have found that the atmosphere height approximately follows the following relation
  \begin{equation}
    h = 10^{8.74} \frac{T_{eff}}{g} \label{atmheight}    
  \end{equation}
  where $h$ is the height in cm of the atmosphere from 1 kbar 
  (approximately the depth where the radiative/convective zone boundary lies)
  to 1 mbar (where the planet becomes optically thin), $g$ is the planet's surface gravity (cgs), and $T_{\rm eff}$ 
  is the effective temperature in Kelvin.  Taking into account this atmosphere height is significant when the planet has
  low gravity or high effective temperature.  In \citet{Fortney07a}, the radii at 1 bar were presented.
\end{enumerate}

The orbital-tidal evolution model is described in detail by \citet{Jackson08,Jackson09} and references therein.
The equations used in this work are
\begin{eqnarray}
  \frac{1}{a} \frac{da}{dt} &=& -\left[ \frac{63 \sqrt{G M_{\mathrm{s}}^3} R_p^5}{2 Q_{\mathrm{p}}' M_p} e^2 \right.\nonumber \\
    &&+ \left.\frac{9 \sqrt{G/M_{\mathrm{s}}} R_{\mathrm{s}}^5 M_p}{2  Q_{\mathrm{s}}'} \left(1 + \frac{57}{4} e^2\right)\right] a^{-13/2} \\
  \frac{1}{e} \frac{de}{dt} &=& -\left[ \frac{63 \sqrt{G M_{\mathrm{s}}^3} R_{\mathrm{p}}^5}{4 Q_{\mathrm{p}}' M_p} + 
    \frac{225 \sqrt{G/M_{\mathrm{s}}} R_{\textrm{s}}^5 M_p}{16 Q_{\mathrm{s}}'} \right] a^{-13/2}\textrm{~~~~} \\
  P_t &=& \frac{63}{4} \left[ (G M_{\mathrm{s}})^{3/2}  \left(\frac{M_{\mathrm{s}} R_{\mathrm{p}}^5 e^2}{Q_{\mathrm{p}}'}\right) \right] a^{-15/2}
\end{eqnarray}
where $a$ is the semi-major axis, $e$ is the eccentricity, and $P_t$ is the
tidal power deposited into the planet.  
This model attempts to describe tidal heating only by orbit circularization and ignores
other forms of tides such as spin synchronization or obliquity tides, which are not
believed to be as important.
This model assumes that the star is rotating slowly relative 
to the orbit of the planet and is second order in eccentricity.  
%% ---ADDED in response to referee's point
Therefore the evolution histories that include periods when the orbit has high eccentricity should
be regarded with caution.
Because there is a lot of other uncertainty with regard to tidal theory, we choose to use this simple model
instead of more complex models such as \citet{Wisdom08}.
%%
%While there are higher-order eccentricity models such
%as \citep{Wisdom08}, given the very large uncertainties in the tidal $Q$ parameter of giant planets, 
%for simplicity we use this simpler classical model.  
For at least 1 of the 45 systems, HAT-P-2, the planet-star system may be able to achieve a double tidally locked equilibrium state 
(star is tidally locked to the planet and the planet is tidally locked to the star) as shown by \ct{Levrard09}; in this system
it is not a good assumption that the star is rotating slower than the period of the orbit.  However, 
\ct{Levrard09} find that this assumption is valid for most stars.
We find that tidal heating is largest where $e$ is not particularly large ($\lesssim 0.4$ falling towards zero) 
so this theory suffices for our purposes.
\qp\ is the tidal $Q$ parameter of
the planet and \qs\ is the tidal $Q$ parameter of the star.  In this work
we have predominantly investigated cases when $Q_{\mathrm{p}}' = Q_{\mathrm{s}}' = 10^{5}$ as well as the case
of $Q_{\mathrm{p}}' = 10^{6.5}$, $Q_{\mathrm{s}}' = 10^5$.  Since the $Q$ value is in principle a function of the driving frequency \citep{OgilvieLin04}, 
amplitude of the distortion, and internal structure of the body, 
the $Q$ value for close-in extra solar giant planets is potentially not equal to the $Q$ value for Jupiter.  
%The value for \qp\ is typically estimated by constraining tidal evolution models for
%the Galilean satellite system \citep{Yoder81}.  The $Q$ value has also been estimated 
%by modeling the dissipation inside the planet in \citep{Goldreich1977} and \citep{OgilvieLin04}.
If the $Q$ value is a very ``spiky'' function of the driving frequency, then the system
might spend a lot of time in a state where the tidal effects are occurring at a slow rate and
quickly pass through states where tidal effects are rapid.  
The stellar $Q$ value is typically estimated through the observed circularization of
binary stars orbits, but has also been estimated by modeling the dissipation inside of a star \citep{OgilvieLin07}.

We assume that the tidal power is uniformly deposited into the envelope
of the planet.  The net energy loss is given by the following equation:
\begin{equation}
  (L - P_t) \Delta t = \int T \Delta S dm.
%                        + \Delta \int_{\textrm{planet}} \frac{G m dm}{r}
\end{equation}
where $L$ is the luminosity at the planet's surface, $\Delta t$ is some small nonzero time step, and $S$ is the specific entropy.  
If $P_t > L$, then the planet's envelope will be increasing 
in entropy and the planet's radius will increase.  More typically, $P_t < L$ and
the planet's entropy is decreasing and thus the planet is contracting.  
The power ratio $P_t / L$ is a useful measure of how important tidal effects are.
It clearly indicates whether there is a net energy input (ratio larger than unity)
or net energy loss (ratio smaller than unity).

For a given radius, assumed core size and average incident flux of the planet, $\dot{R_p} \propto -L_{net}$.  Therefore, 
if we calculate $\dot{R_{NH}}$, the radius contraction rate when there is 
no internal heat source, we can use the following relationship to calculate $\dot{R_p}$ when there is an assumed $P_t$ 
tidal heating (or an input power of another source).
\begin{equation}
  \frac{\dot{R}}{\dot{R_{NH}}} = \frac{L - P_t}{L} \label{rdot}
\end{equation}

Due to tidal migration the incident flux upon the planet increases with time.
Based on the planet's incident flux at a given time, we interpolate in the 4 grids 
which include the incident flux level from the Sun at 0.02, 0.045, 0.1, and 1 AU.  
Here we neglect the more minor effect that parent star spectra can differ somewhat from that of the Sun.

In order to examine all the plausible evolutionary tracks for each 
of the 45 transiting planets studied, we modeled their 
thermal evolution over a range of
\begin{enumerate}
\item
  initial semi-major axis: the observed semi-major axis
  to five times the observed value.
\item
  initial eccentricity: from 0 to 0.8.
\item
  core mass: 0, 10 \me, 30 \me, 100 \me.  For very massive planets we also consider
  core masses of 300 and 1000 \me.  
  Except for GJ 436b, HAT-P-11b, and HD 149026b, the core was required to be at most 70\% of the mass of the planet.  
  For GJ 436b, we sample up to 21 \me and for HAT-P-12, we sample up to 23 \me.
\end{enumerate}

%%% new on stopping condition: begin
Each of these possible evolution histories were run until either a) the time reached
14 Gyr, b) the entropy of the envelope became larger or smaller than the range of entropy values in the grid 
of hydrostatic equilibrium structures, or c) the planet reaches a small orbital distance $\sim R_s$ (realistically, the planet
would be disrupted before this stage, but in this work we do not model the mass loss process).
%% end

For each run, we searched the evolution history during the estimated system age range
for times when the orbital parameters were also within their observed range.  If this occurred we then recorded the transit radius during these times and compared the range of 
achieved values to observed values.
In situations where a good estimate on the age is not available, we searched within
1 to 5 Gyr.  
When a secondary eclipse constraint on the eccentricity is not available 
we assume that the eccentricity value is $0.025 \pm 0.025$ (i.e. the likely range is between 0 and 0.05).
In cases where the eccentricity is observed to be consistent with zero from a secondary eclipse, we assume that the eccentricity 
value is $0.005 \pm 0.005$ (ie. the likely range is between 0 and 0.01).
We use the observed semi-major axis and error.
We then search for instances of evolution histories during the possible age range that have an 
error-normalized distance less than 3 to the observed value.
This distance is defined as
\begin{equation}
  \sqrt{(a_i-a_m)^2/\sigma_a^2 + (e_i - e_m)^2/\sigma_e^2}
\end{equation}
where $a_i$ and $e_i$ are the orbital parameters for the instance of a particular run 
and $a_m$, $\sigma_a$, $e_m$, and $\sigma_e$ are the measured/assumed semi-major axis, semi-major axis sigma, eccentricity, and eccentricity sigma.
Planet orbital parameters, transit radii, 
and stellar parameters are from F.~Pont's website at http://www.inscience.ch/transits/ and 
The Extrasolar Planets Encyclopedia at http://exoplanet.eu/.

\section{General Examples} \label{gen}
Here we add different components of the model step-by-step, such that each effect
can be appreciated independently.  The two opposing effects of tidal evolution are
late-time heating that is associated with eccentricity damping and more efficient
early-time cooling due to initial semi-major axes that are larger then the present value.
The four cases present are for a 1 $M_J$ planet orbiting a 1 $M_{\odot}$ star at 0.05 $AU$.  In each of 
these cases we assume that the planet has a 10 \me\ core.

\begin{figure}
  \begin{center}
  \includegraphics[width=0.94\linewidth]{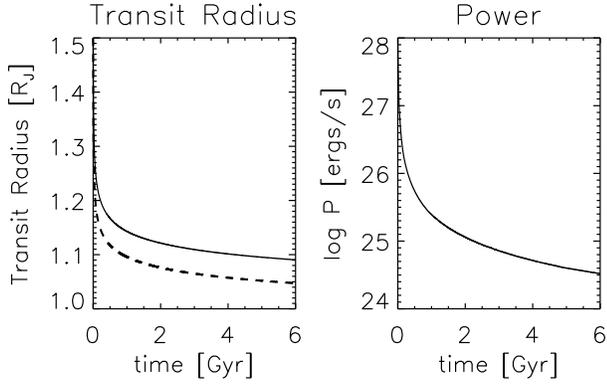}
  \caption{Radius and intrinsic planet luminosity evolution for a 1 $M_J$ planet at 0.05 AU around a 1 $M_\odot$ star without any tidal effects.  
    In the left panel, the dashed line is the radius at 1 kbar, near the convective/radiative boundary at gigayear ages.
    The solid line is the radius where the atmosphere reaches 1 mbar - approximately
    the radius that would be observed in transit.}
  \label{stationary}
  \end{center}
\end{figure}

Case 1: no tidal effects, Figure \ref{stationary}.  
In the left panel,
the solid line is the planet transit radius and the dot-dashed line is the radius at 1 kbar (near the convective-radiative boundary).  
In the right panel, the intrinsic planet luminosity is plotted as a function of time.  As the planet contracts the luminosity of the planet
significantly decreases.  Without an internal heat source or semi-major axis evolution the planet's radius monotonically decreases with time.

\begin{figure}
  \begin{center}
  \includegraphics[width=0.94\linewidth]{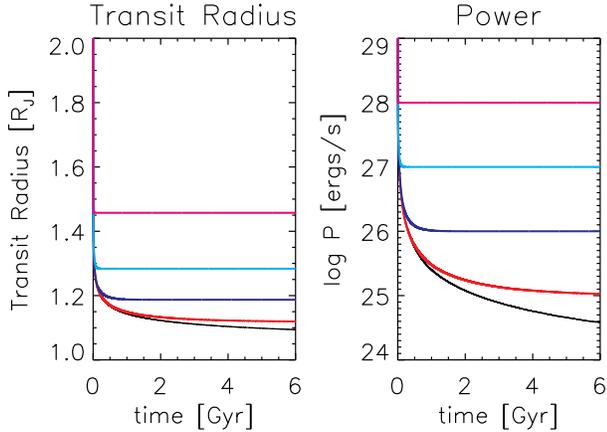}
  \caption{Similar to Figure \ref{stationary}, but with various constant heating applied in the interior of the planet.  Moving from bottom to top, 
    the constant heating rates are $10^{24}$, $10^{25}$, $10^{26}$, $10^{27}$ and $10^{28}$ erg s$^{-1}$.}
  \label{constheating}
  \end{center}
\end{figure}

Case 2: no orbital evolution, constant interior heating, in Figure \ref{constheating}.  In this case the net output power is the difference between
the intrinsic luminosity and a constant interior heating source of unspecified origin.  In these evolution runs, the planet stops contracting
when the intrinsic luminosity is equal to the constant heating source.  This is equivalent to when the ratio between the input power and the luminosity
of the planet is equal to unity.
The upper 3 evolution tracks (purple, cyan, and blue) all reach
an equilibrium between the interior heating and luminosity of the planet within 2 Gyr, but the evolution runs with lower input power do not reach
an equilibrium state in the 6 Gyr plotted.  
As expected, when there is more input power, the equilibrium radius is larger.  
In practice, the input power through tides or other processes will not be constant over gigayears, 
but a planet may be inflated to a radius such that it is in a temporary equilibrium state.

\begin{figure}
  \begin{center}
  \includegraphics[width=0.94\linewidth]{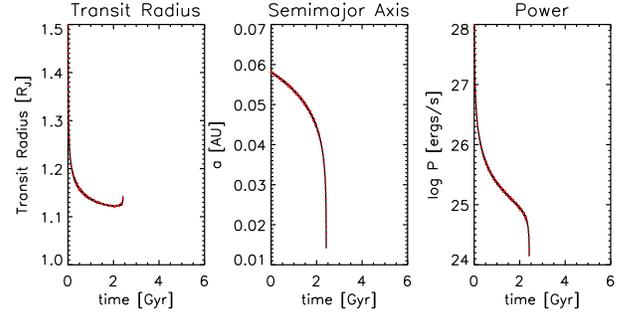}
  \caption{Planet thermal evolution with orbit evolution, but without tidal heating.  Transit radius, semi-major axis and the planet's
  intrinsic luminosity are plotted from left to right.  $Q_{\mathrm{p}}' = 10^5$ and $Q_{\mathrm{p}}' = 10^{6.5}$ 
  cases are plotted in black and red respectively.}
  \label{NHev}
  \end{center}
\end{figure}

Case 3: tidal orbital evolution, but without tidal heating, Figure \ref{NHev}.  This case demonstrates how the orbital evolution
due to tides effects the thermal evolution of the planet.
Here we plot both the $Q_{\mathrm{p}}' = 10^5$ (tidal effects on the planet occur faster) and $Q_{\mathrm{p}}' = 10^{6.5}$ (tidal effects on the planet occur slower)
cases with $Q_{\mathrm{s}}' = 10^5$ in black and red respectively.
These curves exactly track each other because the tides on the planet do not significantly contribute to
the migration when the eccentricity is small (here $e=0$).  
When comparing Figure \ref{stationary} to Figure \ref{NHev}, notice that in the second case, the power drops off 
more rapidly as the semi-major axis decreases.  This is due to the increase in insolation by the parent star, which deepens the atmospheric radiative zone, lessening transport of energy from the interior \citep[e.g.][]{Guillot96}.  
Another result of moving the planet closer to the star is that there is an up-tick in the transit radius.  This is
due only to an increase in the effective temperature, which increases the atmosphere height.
The semi-major axis evolution accelerates as the planet moves inward due to the tidal migration rate's strong dependence on semi-major axis.

\begin{figure}
  \begin{center}
  \includegraphics[width=0.94\linewidth]{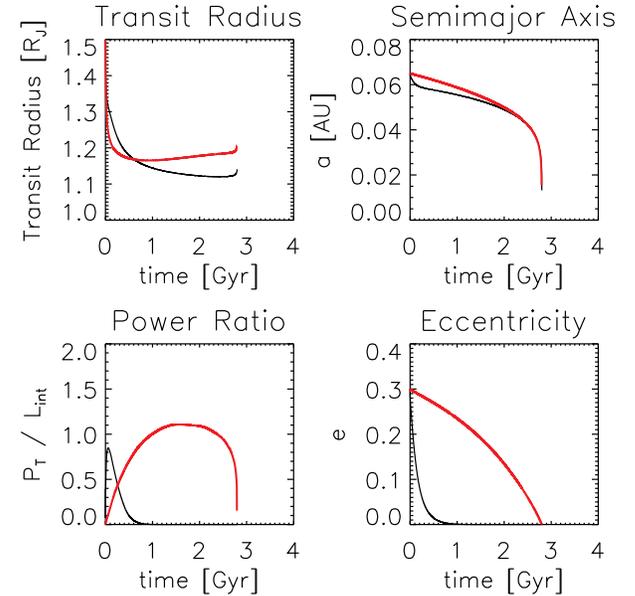}
  \caption{Coupled planet thermal evolution and orbital evolution.  $Q_{\mathrm{p}}'=10^5$ and $Q_{\mathrm{p}}'=10^{6.5}$ cases both with
    $Q_{\mathrm{s}}'=10^5$ are plotted in black and red respectively.  We plot, the radius evolution in the upper left, semi-major axis evolution in the
  upper right, ratio between tidal heating and intrinsic planet luminosity in the lower left, and eccentricity in the lower right.  
  }
  \label{tidalev}
  \end{center}
\end{figure}

Case 4: tidal orbital evolution and tidal heating, Figure \ref{tidalev}.  
We now put both the orbital evolution and corresponding tidal heating together.  
Black is the $Q_{\mathrm{p}}' = 10^5$ case and red is $Q_{\mathrm{p}}' = 10^{6.5}$ case.  
Notice that in the low $Q_{\mathrm{p}}'$ case, the planet
circularizes quickly and tidal heating becomes less important.   
In the high $Q_{\mathrm{p}}'$ case, the planet is still undergoing circularization and significant tidal heating at late times.
As a result, the radius in the high \qp\ case (slower rate of tidal effects in planet) can be larger than the low \qp\ case 
(faster rate of tidal effects in planet) at late times.  Both trials start out with fairly modest eccentricity ($e=0.3$).

\begin{figure}
  \begin{center}
  \includegraphics[width=0.94\linewidth]{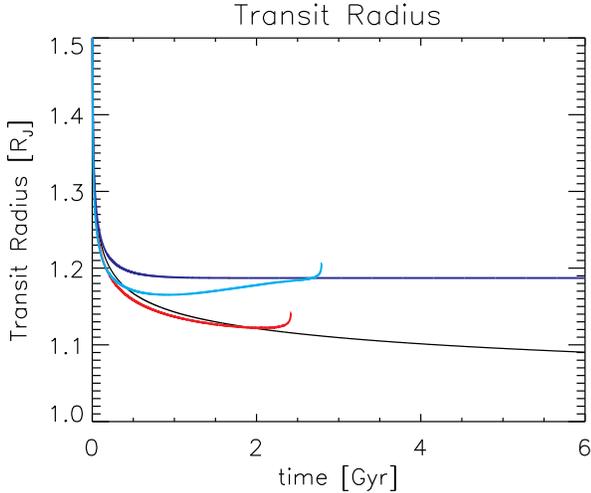}
  \caption{Radius evolution in different cases.  Cases 1, 2, 3, 4 (see text) are plotted in black, blue, red, and cyan.}
  \label{allev}
  \end{center}
\end{figure}

In Figure \ref{allev}, we compare the radius evolution in all four of these cases:  Case 1 (no tidal effects, black), 
Case 2 (no orbital evolution, constant heating, blue), Case 3 (tidal orbital evolution, but not tidal heating, red), 
and Case 4 (full tidal evolution model, cyan).
The cases with tidal evolution are plotted for the high \qp\ case.  Clearly, when tidal heating is included (cyan or blue), 
it can result in a radius larger than achieved without
including tidal heating (red or black).  Since tidal heating is a time-varying quantity, 
the planet's radius when tidal heating will not be as simple as in Case 2.  Generally,
the planet will experience significant tidal heating when the orbit is being circularized.  
At this time, the radius will increase, but after this time the radius of the planet
will contract again.  Also, because the planets in Case 3 (red) start at larger orbital distance than that of Case 1 (black), 
the radius contracts marginally faster when the planet
is at larger semi-major axis.  This is why the red line is lower than the black line before 2 Gyr.  
After this point, the transit radius increases in the red line case because
the planet has moved close to the star, the effective temperature of the planet increases, and the atmosphere height also increases.  

\begin{figure}
  \begin{center}
  \includegraphics[width=0.94\linewidth]{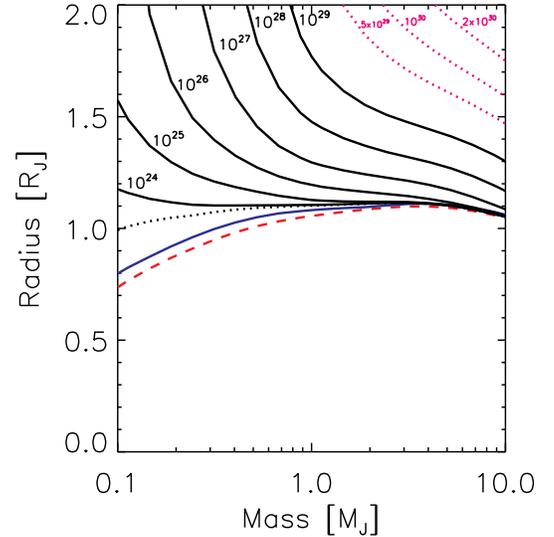}
  \caption{
    The dotted black line is the transit radius without any
    internal heating as a function of mass assuming a 10 \me\ core.  In these models, we
    hold the planet at 0.05 AU around a 1 Solar Mass star.
    The dashed red line is the 1 kbar radius---near the convective/radiative zone boundary.
    The blue line is the relation from \citet{Fortney07a}.  
    The solid black lines
    are the radius one would find if there were a constant heating source (values between $10^{24}$ and $10^{29}$ erg s$^{-1}$). 
    The pink dotted lines were calculated in the same way, but required extrapolation (quadratic) off of the 
    grid of atmosphere models.
  }
  \label{radofm}
  \end{center}
\end{figure}

To examine how different levels of internal heating affect the radius of the planet, we plot the planet radius
after 5 Gyr as a function of mass in Figure \ref{radofm}.  Again, these models assume a 10 \me\ core, at a orbital distance of 0.05 AU around
a 1 Solar Mass star.
In this figure, the black dotted line is the prediction of the thermal evolution model without tidal heating.  The red
dashed line is the base of the atmosphere at 1 kbar.  Clearly, the height of the atmosphere is much larger for smaller planets
due to their smaller gravities.  
The solid blue line is the radius relation from \citep{Fortney07a}.
The solid black lines are the radius of the planet given a constant heating rate after 
5 Gyr of evolution.  
The pink dotted curves are constructed in the same manner as the solid black curves, but required
extrapolation (here, quadratic) off of the calculated atmosphere grid.
At this point in time, most of these planets have reached an equilibrium state where
an equal amount of internal heating is balanced by the planet's intrinsic luminosity.  Clearly, the effect on the radius
for a given heating is larger for smaller mass planets.  

\section{Results} %% GO THROUGH AGAIN, CHECK Brian's changes
\subsection{Specific Systems}
\begin{figure}
  \begin{center}
  \includegraphics[width=0.94\linewidth]{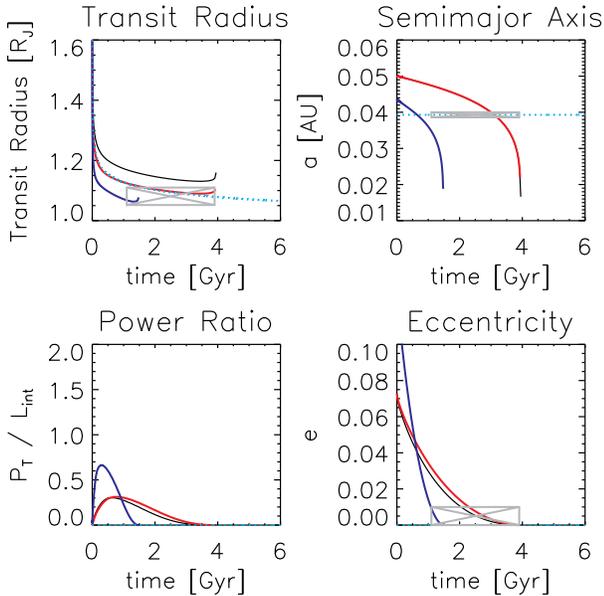}
  \caption{Possible tidal/thermal evolution tracks for the planet
    around the star TrES-1.  Black: no core.  Red: 10 \me\ core.
    Blue: 30 \me\ core.  
    Cyan dotted: 10 \me\ core evolution history without
    tidal effects.  
    This is a 0.76 \mj\ planet orbiting a 0.89 $M_{\odot}$ star.
    Upper left panel: transit radius evolution.
    Upper right panel: semi-major axis evolution.
    Lower left panel: ratio between tidal power injected into the planet 
    and intrinsic planet luminosity.
    Lower right panel: eccentricity evolution.
    Observed semi-major axis, eccentricity and
    observed radius are plotted in their respective panels.  These
    evolution tracks were selected to have orbital parameters that
    agree with the observed values.  $Q_{\mathrm{p}}' = 10^{6.5}$, $Q_{\mathrm{s}}' = 10^{5}$.
  }  
  \label{TrES1evol}
  \end{center}
\end{figure}

\begin{figure}
  \includegraphics[width=0.94\linewidth]{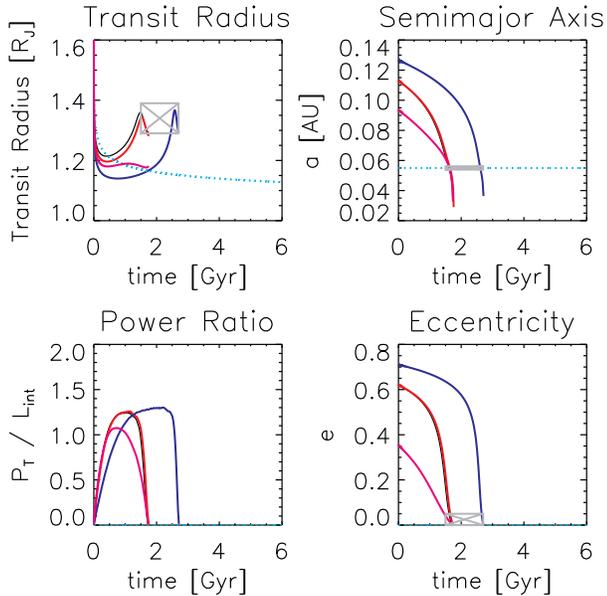}
  \caption{Possible tidal/thermal evolution tracks for the planet
    around the star XO-4. 
    This is a 1.72 \mj\ planet orbiting a 1.32 $M_{\odot}$ star.
    Black: no core.  Red: 10 \me\ core.
    Blue: 30 \me\ core.  Magenta: 30 \me\ core with a low initial eccentricity.
    Cyan dotted: 10 \me\ evolution history without tidal effects.
    Panels are analogous to Figure \ref{TrES1evol}.
%    Upper left panel: transit radius evolution.
%    Upper right panel: semi-major axis evolution.
%    Lower left panel: tidal power injected into the planet (solid)
%    and intrinsic planet luminosity (dashed).
%    Lower right panel: eccentricity evolution.
%    Observed semi-major axis and observed radius are plotted in their respective panels.  
    The eccentricity that is marked in the lower right panel is our
    assumed possible range (0 to 0.05).
    These evolution tracks were selected to have orbital parameters that
    agree with the observed values.  $Q_{\mathrm{p}}' = Q_{\mathrm{s}}' = 10^{5}$.
    Notice that the tidal models initially have smaller radii than the non-tidal model
    because the tidal models are able to more efficiently cool at early times due to their larger semi-major axis.
  }
  \label{XO4evol}
\end{figure}

\begin{figure}
  \begin{center}
  \subfloat[] {\label{XO4gridA} \includegraphics[width=0.9\linewidth]{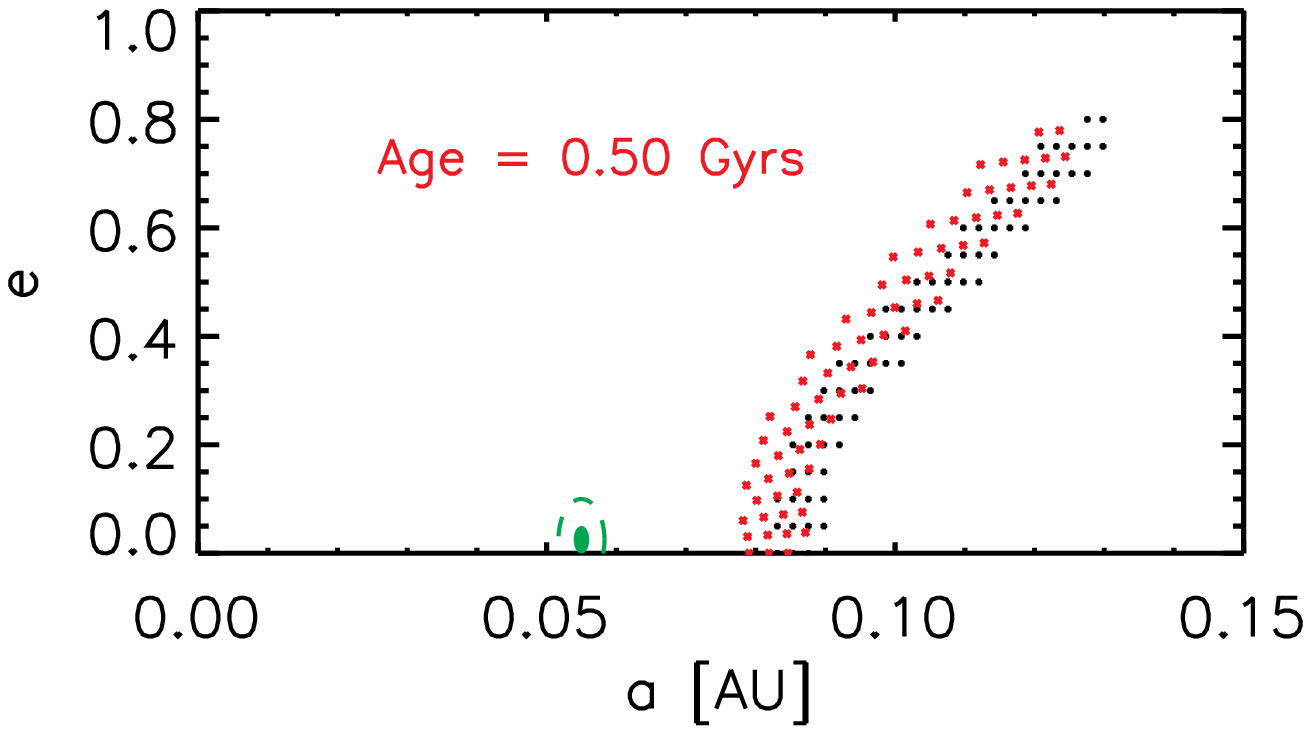}} \\
  \subfloat[] {\label{XO4gridB} \includegraphics[width=0.9\linewidth]{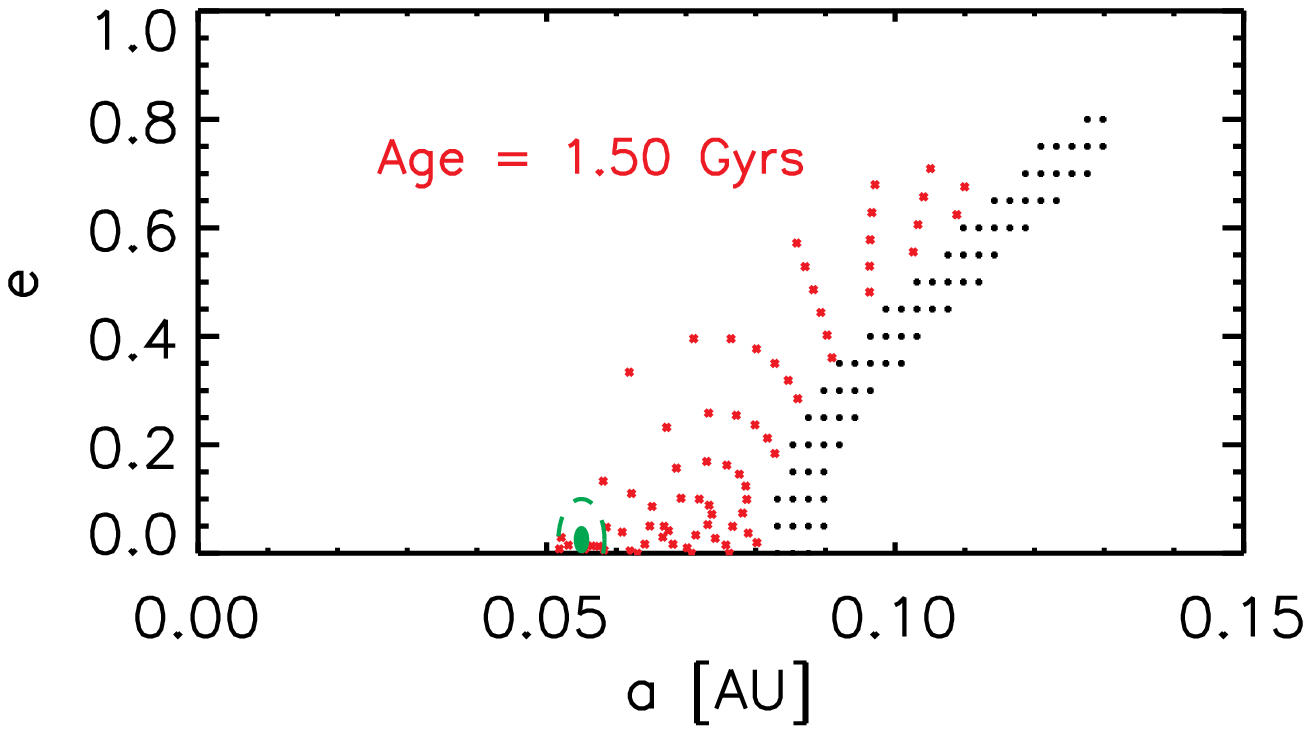}} \\
  \subfloat[] {\label{XO4gridC} \includegraphics[width=0.9\linewidth]{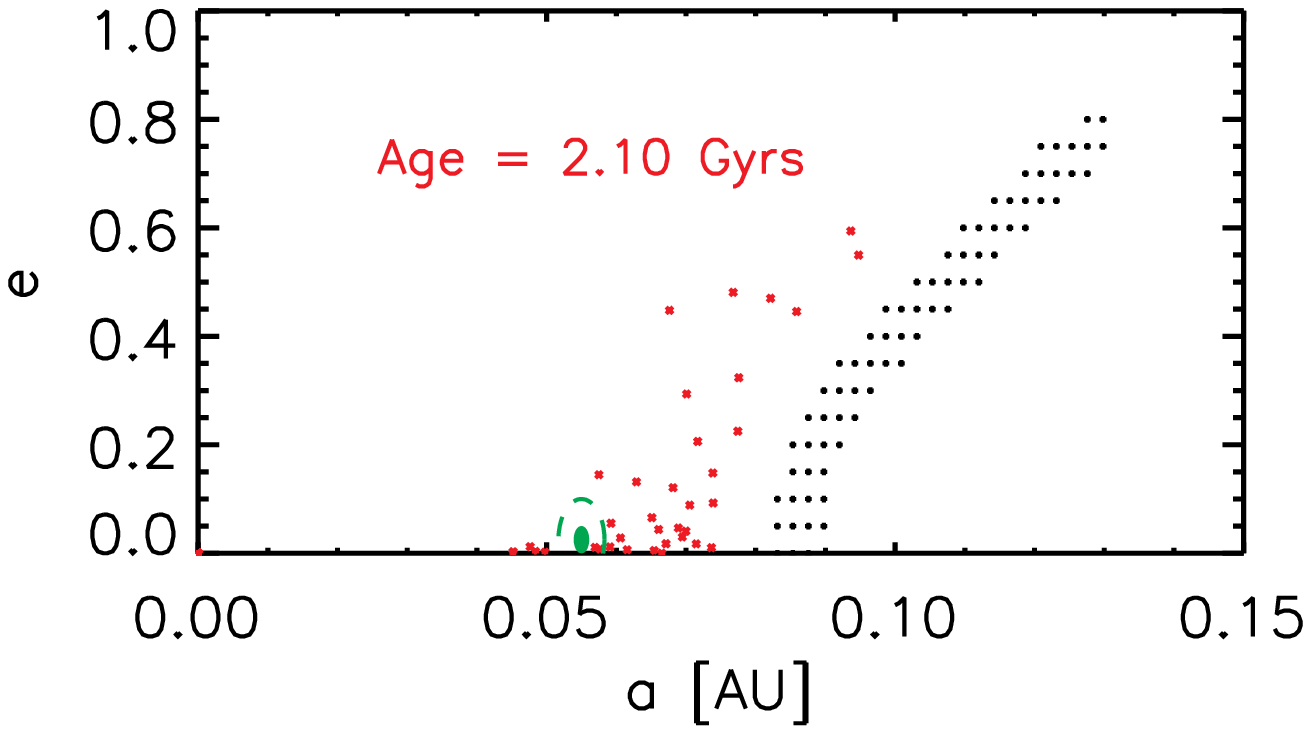}} 
  \caption{
    Grid of evolution histories for XO-4b that were found to be consistent with the orbital parameters at a later time. 
    These histories are not required to also have a radius value that is consistent with the observed value.
    These evolution runs assume a core size of 10 \me\ , \qp\ $=10^5$, and \qs\ $=10^5$.
    This serves as a sample for the type of calculation that was performed for every planet.
    Black: original orbital parameters of each run.  Red: orbital parameters at a later marked time (0.5 Gyr, 1.5 Gyr, and 2.1 Gyr).
    The filled green circle is the 1 $\sigma$ zone, while the dashed region is the 3 $\sigma$ zone.
  }
  \label{XO4grid}
  \end{center}
\end{figure}

\begin{figure}
  \includegraphics[width=0.94\linewidth]{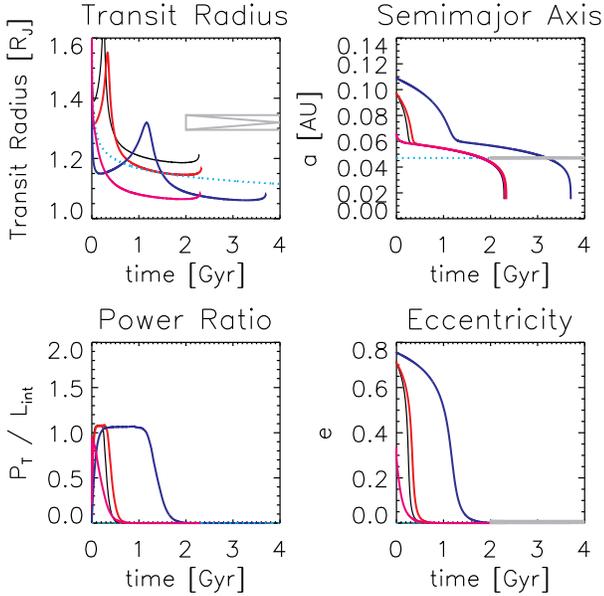}
  \caption{Possible tidal/thermal evolution tracks for the planet
    around the star HD 209458.  
    This is a 0.657 \mj\ planet orbiting
    a 1.101 $M_{\odot}$ star.  The planet has a radius of 1.32 \rj\ and 
    an observed eccentricity of zero.
    Black: no core.  Red: 10 \me\ core.
    Blue: 30 \me\ core. Purple: 30 \me\ core with low initial eccentricity.
    Cyan dotted: 10 \me\ core evolution model without tidal effects.
    Panels are analogous to Figure \ref{TrES1evol}.
    $Q_{\mathrm{p}}' = Q_{\mathrm{s}}' = 10^{5}$.       
%    Upper left panel: transit radius evolution.
%    Upper right panel: semi-major axis evolution.
%    Lower left panel: tidal power injected into the planet (solid)
%    and intrinsic planet luminosity (dashed).
%    Lower right panel: eccentricity evolution.
%    Observed semi-major axis, eccentricity and
%    observed radius are plotted in their respective panels.  
%    This case demonstrates that in order to observe the eccentricity
%    to be consistent with zero (between 0 and 0.01), it is not possible in this case to also
%    observe an inflated radius.  
%    It also demonstrates that tidal heating can provide sufficient power to inflate the radius, but
%    that this observation could only occur when the eccentricity is nonzero.  
%    Notice that the semi-major axis evolution appears to have a knee.  This occurs because early
%    in these evolution histories, the tides on the planet are important when the eccentricity is large.  
%    After the orbit has circularized, the tides and the star are the dominant cause of orbital migration.
  }
  \label{HD209458evol}
\end{figure}

\begin{figure}
  \includegraphics[width=0.94\linewidth]{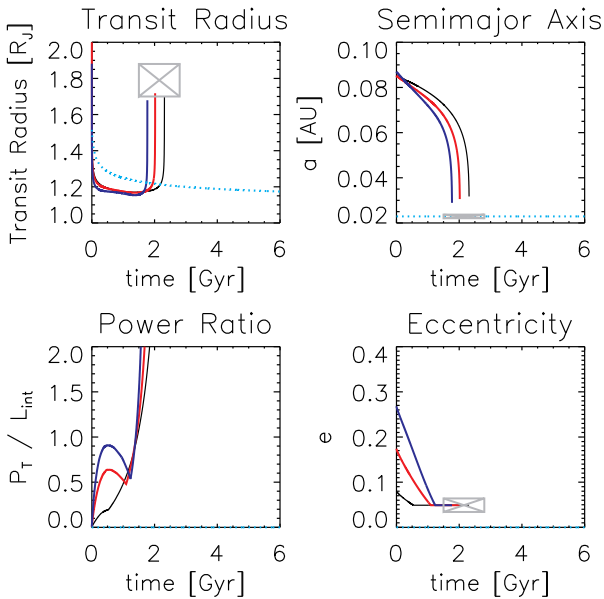} %SampEvWASP-12-Qp5Qs5_PlusMinERun.ps}
  \caption{
    Possible tidal/thermal evolution for WASP-12b.  This is a 1.41 \mj\ planet
    orbiting a 1.35 $M_{\odot}$ star.  The planet has a very large observed transit radius
    of 1.79 \rj\ and an eccentricity of 0.05.  In these evolution histories, we impose an eccentricity
    floor mimicking the effects of an eccentricity driving force.
    Black: no core.  Red: 10 \me\ core.
    Blue: 30 \me\ core.  
    Cyan dotted: 10 \me\ core evolution history without tidal effects.
    Panels are analogous to Figure \ref{TrES1evol}.
    \qp\ $=10^5$ and \qs\ $=10^5$.
%    Upper left panel: transit radius evolution.
%    Upper right panel: semi-major axis evolution.
%    Lower left panel: tidal power injected into the planet (solid)
%    and intrinsic planet luminosity (dashed).
%    Lower right panel: eccentricity evolution.
%    Observed radius, semi-major axis and eccentricity are plotted in their
%    respective panels.  
%    Notice that the observed radius is achieved by the case with
%    by the minimum initial eccentricity, but a floor place in place at the current time.  Tidal
%    heating becomes more important enough to significantly increase the radius of the planet as the semi-major axis decreases.
%    In this run, the planets radius rapidly continues to increase beyond the observed value.  This calculation
%    is meant to show that if the eccentricity is maintained at a nonzero value, then it can be possible to
%    significantly inflate the planet's radius.  
  }
  \label{WASP12evol}
\end{figure}

While we have computed the evolution history of 45 systems, here we show representative calculations for particular samples of planets.  
These are TrES-1b, XO-4b, HD 209458b, and WASP-12b, 
and are shown in Figures \ref{TrES1evol}, \ref{XO4evol}, \ref{HD209458evol}, and \ref{WASP12evol} respectively.  
These four cases demonstrate qualitatively different cases.  TrES-1b is a circularized planet with a ``normal'' radius value.
XO-4b, HD 209458b, and WASP-12b are large-radii planets with a small relatively unconstrained
eccentricity, zero eccentricity, and a nonzero value, respectively.
In Figures \ref{TrES1evol} - \ref{WASP12evol},
the transit radius evolution is plotted in the upper left panel,
the semi-major axis evolution is plotted in the upper right panel,
the ratio between the tidal power and luminosity is plotted in the lower left panel,
and the eccentricity evolution is plotted in the lower right panel.
The observed semi-major axis, eccentricity, and transit
radius are plotted on each of the respective panels.
The power ratio, tidal power to luminosity, describes how
important tidal effects are to the energy flow of the planet.  
When this ratio is somewhat smaller than unity, tidal heating is relatively un-important for the thermal evolution of the planet
and when this ratio reaches or surpasses unity, tidal heating plays a more significant role in the thermal evolution.
In each of these figures, a set of runs were selected such
that the orbital parameters and transit radius are closest to the observed values.

TrES-1b is a transiting hot-jupiter planet with zero or small eccentricity and a typical radius observation.  
The system is composed of a 0.76 \mj\ planet orbiting a 0.89 $M_{\odot}$ star with a 0.04 AU semi-major axis.
Tidal heating is not necessary to invoke to explain this system; we demonstrate that this tidal model
can still explain these kinds of modest radius systems.
Possible evolution histories with tidal effects are shown in Figure \ref{TrES1evol}.
These possible histories are selected such that their orbital parameters at the current age agree with
the observed values and the transit radius that is close to the observed value.
We show various core sizes in different colors: black for zero core,
red for a 10 \me\ core, and blue for a 30 \me\ core.  The cyan dotted line is the evolution history of a non-tidal
thermal evolution model with a 10 \me\ core.  Notice the radius evolution of the non-tidal model doesn't differ significantly
from the radius evolution of the corresponding 10 \me\ (red) tidal model.
In these possible evolution histories with tidal effects, the initial eccentricity is relatively small and tidal heating doesn't dominate
the energy flux budget (in the lower left panel, the power ratio is always less than 1).
However, the orbit decays significantly due to tides raised on the star by the planet, which continues even at $e=0$.  These tides cause
these planet to migrate from an initial semi-major axis of 0.05 AU to 0.04 AU with the assumed $Q_{\mathrm{s}}' = 10^5$.  
%This tidal-orbital evolution allows the planet to more efficiently cool off at early times, 
%compared to the prediction from a stationary non-tidal model.  
Figure \ref{TrES1evol} demonstrates that this model easily explains the radius of TrES-1b with a core 
between 10 \me\ and 30 \me.

There is a slight upturn in radius just before an age of 4 Gyr.  
This is due to the heating of the planet's atmosphere at very small semi-major axis, and is \emph{not} due to tidal power.  
As the planet reaches smaller orbital distances the incident flux it intercepts increases dramatically, 
leading to an enlarged atmospheric extension, and greater transit radius.  This feature is also present in the 
recent paper by \citet{IbguiBurrows2009}.
The tracks end when we stop following the evolution, with the assumption that the planet is disrupted or collides with the parent star.  
This is merely the first of many evolution tracks that we present with the end state being the disruption of the planet.  
This finding is essentially quite similar to that of \ct{Levrard09} who find that all of the known transiting planets, 
save HAT-P-2b, will eventually collide with their parent stars.  Robust observational evidence for this mechanism was recently detailed by \ct{Jackson09}.

XO-4b is an inflated planet where the eccentricity has not been well constrained, due to sparse radial velocity sampling \cp{McCullough08}.  
In these cases we search for instances over the evolution histories where the eccentricity is between 0 to 0.05, 
because we assume that a larger value would have been clearly noticed in radial velocity data.
With this eccentricity constraint we show in Figure \ref{XO4evol} that there is a narrow period of time
when we can explain the inflated state with a recent circularization of the orbit that has deposited
energy into the interior of the planet.  The evolution curves shown here are for tidal parameters $Q_{\mathrm{p}}'=10^5$ and
$Q_{\mathrm{s}}'=10^5$; in the higher $Q_{\mathrm{p}}'$ case, the radius evolution curves do not agree with the observed value.
In Figure \ref{XO4evol}, we show black, red, and blue curves for evolution runs with no core, 10 \me\ core, and
30 \me\ core respectively.  The pink curve is an evolution
history for low initial eccentricity with a 30 \me\ core.  Again, the cyan curve is a no-tidal evolution history with
10 \me\ core.
Since tidal power is deposited mainly when the planet is being circularized, {\it high initial eccentricity orbits are required for
these planets to experience significant later tidal inflation}.  Another interesting feature of this plot, is that when comparing
the radius of the runs for different cores at any given time, we find that the radius is not always monotonically decreasing
with core size.  This shows that uncertain past orbital-tidal history can lead to uncertainly in derived structural parameters such as the core mass.

As an example of the kind of calculation that was performed for every planet, in Figure \ref{XO4grid} we show snapshots of the orbital parameters ($a$ and $e$) of the ensemble of systems that are at some point consistent
with the observed orbital parameters and age of XO-4b.  
Note that we do not require that the radius simultaneously also agree with the observed radius, but rather
compare the range of possible radius values achieved by the model to the actual observed value.
The black points are the original orbital parameters.  The red points are the 
orbital parameters for one of these runs at a later point in time (0.5 Gyr, 1.5 Gyr, and 2.1 Gyr).  The filled green circle
marks the 1 $\sigma$ observed orbital parameters, while the dashed region is the 3 $\sigma$ zone.  

HD 2094598b is a large-radius planet with eccentricity that has been observed to be very close to zero \cp{Deming05b}.  
The planet is observed to have a radius of 1.32 \rj\ and mass of 0.657 \mj.  
Therefore we require evolution histories where the current eccentricity is $< 0.01$.
Evolution histories for this system are shown in Figure \ref{HD209458evol} with \qp\ $=10^5$ and \qs\  $=10^{5}$.
With these chosen $Q$ values, we find that the planet could have
experienced tidal heating at a previous time, however by the time it has an eccentricity
of 0.01 or less the planet's radius has since deflated below the observed value.  
It is possible to find an evolution histories that agrees with the observations by allowing different $Q$ values, as shown by
\citet{IbguiBurrows2009}.  Although the tidal $Q$ value is not strongly constrained and may even vary depending
on the configuration of the system \citep{OgilvieLin04}, it is our view that it makes the most sense
to fix the $Q$ value close to prior inferred values.
Again, the black, red, and blue curves correspond to no core, 10 \me\ core, and 30 \me\ core
sizes respectively.  
The cyan curve is a non-tidal thermal evolution history for a 10 \me\ core.
In these cases, tidal power is sufficient to inflate the planet's radius to its observed value,
however we do not find evolution histories that also agrees with the other observed parameters---especially the eccentricity.
In the semi-major axis evolution, there is a clear transition knee where the rate of orbital evolution decreases.  The first
phase is due to tidal effects of both the star and planet while the eccentricity is nonzero.  The second phase is mainly due
to tides on the star when the eccentricity is zero.

WASP-12b is a planet with an especially large radius of 1.79 \rj\ with a non-zero eccentricity of 0.05 \cp{Hebb09}.
An interesting property of this system is that the planet is filling at least 80 \% of its Roche lobe by radius \cp{Li09}.
Figure \ref{WASP12evol} shows evolution curves in black, red, and blue for
no core, 10 \me\ core, and 30 \me\ core cases respectively when an eccentricity floor is imposed.  
Also, in cyan is the non-tidal model.
In these tidal cases the tidal power increases in strength as the semi-major axis decays until the 
planet undergoes a rapid expansion.  
%In this case , we rapidly inflate the radius beyond the observed value.
When the semi-major axis gets small enough, the tidal power exceeds the luminosity and
the planet's radius rapidly increases.  This happens both because the incident flux decreases
the intrinsic luminosity of the planet and tidal heating has a strong semi-major axis dependence 
($P_t \sim a^{-15/2}$).
%Our models do not take into account the effects of mass loss by tidal stripping or other exotic processes that may be occurring here.
We do not model the mass loss process, which is likely to occur at late times for systems such as these \citep{Gu03}
This should only be taken as evidence that if there was an eccentricity driving companion similar to mechanisms suggested by \citet{Mardling2007},
then it may be possible to heat this planet to quite large radii.  

%Note that the \citep{Jackson08} tidal evolution model omits higher order eccentricity terms.  Therefore the rate that tidal evolution occurs at high eccentricity may be questionable. However, this does not resolve the issue with the low eccentricity planets that have large radii.  The tidal model should at least be a good estimate the amount of tidal power that is possible for these systems in recent times when the eccentricity was low.

\begin{figure*}
  \subfloat[] {\label{plrad5A} \includegraphics[width=0.8\linewidth]{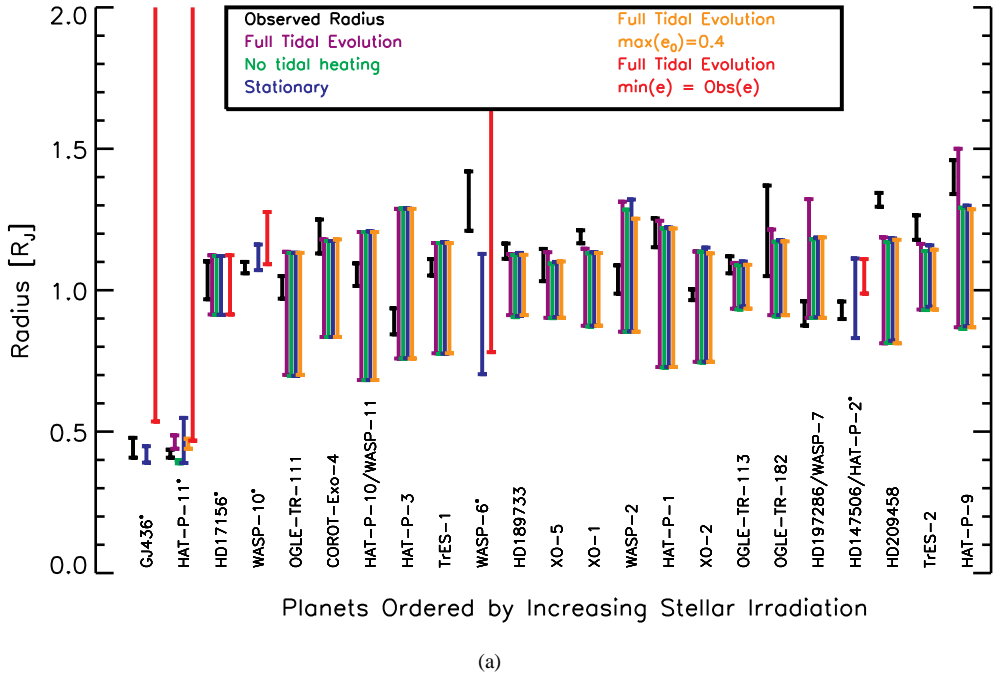}} \\
  \subfloat[] {\label{plrad5B} \includegraphics[width=0.8\linewidth]{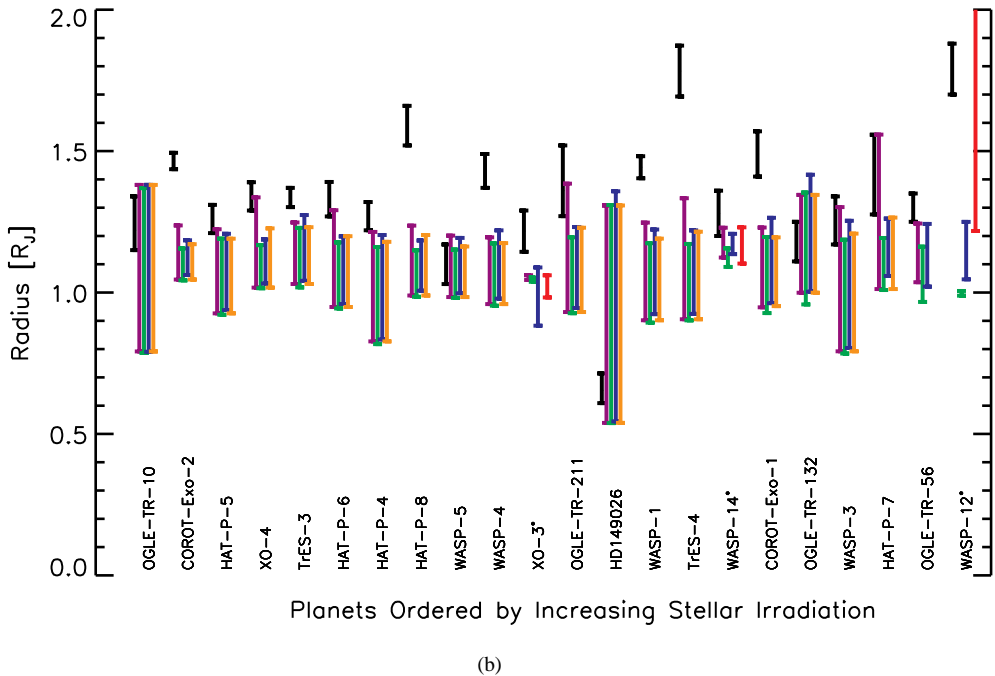}}

  \caption{    
    Observed planet radius (black) compared to a range 
    of achieved model radii (colors) using $Q_{\mathrm{p}}' = 10^{5}$; $Q_{\mathrm{s}}' = 10^{5}$.  
    Planets are ordered by increasing incident flux according to their 
    current observed parameters.  
    Planets are marked with a * if they have nonzero observed eccentricity.
    The range of possible radius values under the full tidal evolution
    model is plotted in purple with initial eccentricity between 0 and 0.8.
    The radius range for a model with tidal-orbital evolution, 
    but without the tidal heating into the interior of the planet is plotted in green.
    The radius range for a standard stationary model without any tidal effects is plotted in blue.  
    The radius range for the full tidal evolution
    model with a maximum initial eccentricity of 0.4 is plotted in orange.  
    In cases where a nonzero eccentricity has been observed, the radius range 
    with an eccentricity floor equal to the observed value is shown in red.
    \label{plrad5}
  }
\end{figure*}

\begin{figure*}
  \subfloat[] {\label{plrad65A} \includegraphics[width=0.8\linewidth]{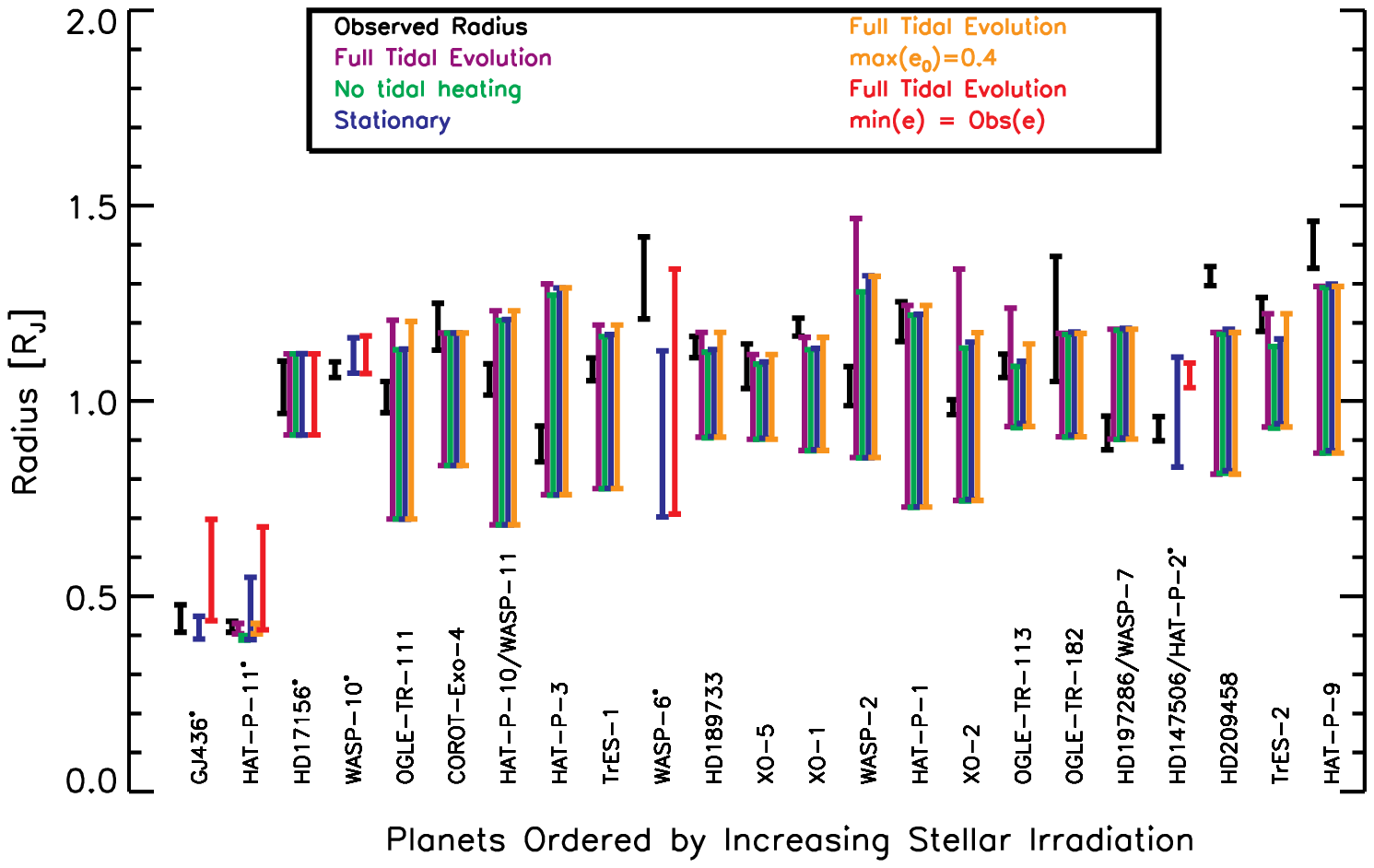}} \\
  \subfloat[] {\label{plrad65B}   \includegraphics[width=0.8\linewidth]{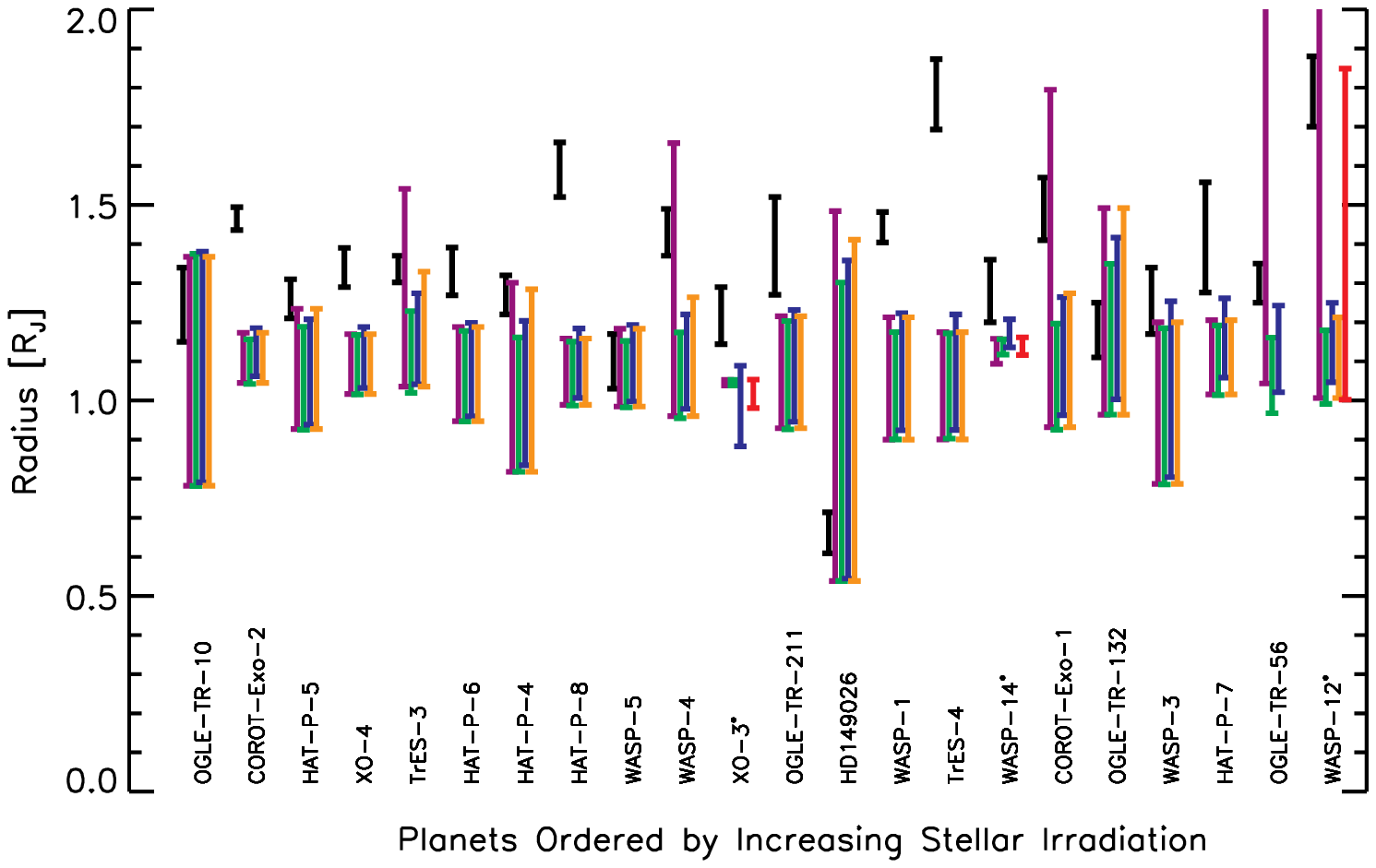}}

  \caption{
    Observed planet radius (black) compared to a range 
    of viable model radii (colors) using $Q_{\mathrm{p}}' = 10^{6.5}$; $Q_{\mathrm{s}}' = 10^{5}$. 
    Qualitatively, we observe the same trends that were observed in Figure \ref{plrad5A} and (\ref{plrad5B}).
    A larger $Q_{\mathrm{p}}'$ value decreases the rate of tidal effects via tides on the planet.  
    Typically the tides on the planet from the star are responsible for circularizing 
    the orbit, while tides on the star from the planet are responsible for decreasing
    the semi-major axis.  In the larger $Q_{\mathrm{p}}'$ case, the tidal circularization can be delayed
    for longer, which can make the possible radius of the planet larger.  On the other hand,
    a larger $Q_{\mathrm{p}}'$ also decreases the power deposited into the planet.  
    %With these parameters,
 %   HD 147506, COROT-Exo-2, XO-4, TrES-4, and WASP-12 were not explainable with tidal heating.
 %   HAT-P9, COROT-Exo-1 and WASP-1 were explainable by tidal heating and not with the stationary
 %   model.  Notice that WASP-12 and XO-4 are no longer explainable via tidal heating with $Q_{\mathrm{p}}'=10^{6.5}$, but
 %   under some assumptions these were explainable with $Q_{\mathrm{p}}'=10^5$.
    \label{plrad65}
  }
\end{figure*}

\subsection{Summary for Suite}
% Paragraph 1: define what each of these are
We have summarized our results for all 45 planetary systems in Figures
\ref{plrad5} and \ref{plrad65} for \qp\ equal to $10^5$ and $10^{6.5}$.  
In these figures, we have plotted
the observed radius range (lower limit to upper limit) in black. 
The achieved radius range under various assumptions is plotted in color.
%The range of achieved radius values for each model is due to multiple factors.
%First, each model is run with multiple core sizes (typically 0-100 \me).
%In some cases, we also run 300 and/or 1000 \me core size.  In all cases except
%GJ 436b, we require that the core be at most 70\% of the planet's mass.
%Second, 
Possible radii are recorded in instances of the evolution histories
when the orbital parameters and age all agree with the observed $a$, $e$, and age values 
(as defined previously, within 3 error-normalized distance units of the observed value).  
The age of each system is often quite uncertain; since the possible radius values are 
sensitive to the age of the system, this is a large source of uncertainty for our results.
For each planet, a range of radius values is plotted for up to five different successful types of models.  
These are models computed as discussed in \S \ref{gen}.
\begin{enumerate}
\item
The full tidal evolution model is shown in purple.   
In this model the initial eccentricity was sampled from 0 to 0.8 and the initial
semi-major axis was sampled from the observed semi-major axis to 5 $\times$
the observed value.  This is case 4 in \S \ref{gen}.
\item
The model with tidal migration but without heating is shown in green.  We perform the same 
search procedure as in the full tidal model.  This model is not meant to be physical, but
to give us an understanding of how tidal orbital migration alone effects the planet's radius.
This is case 3 in \S \ref{gen}.
\item
The ``stationary'' model is shown in blue with all tidal effects turned off.
These are ``standard'' cooling/contraction models, quite similar to those in \ct{Fortney07a}.
These models differ slightly than the models listed in \ct{Fortney07a} in two ways.
First, these models more accurately take into account the height of the atmosphere.
Second, some of these models explore a wider range of core sizes.
%The radius range achieved in this work varies a bit more than listed in \ct{Fortney07a}.  
%This is explained by three effects.
%First, in this work the height of the atmosphere is being taken into account.  
%Second, this work searches for the possible radius values over the whole range of possible system ages.
%Third, the radius range that is shown is over a variety of core sizes.  
This is case 1 in \S \ref{gen}.
\item
For planets whose current observed eccentricity is less than 0.4, the full tidal evolution 
with an maximum initial eccentricity of 0.4 is plotted in orange.
Because tidal heating in the planet is directly connected to eccentricity damping, these
runs serve as a demonstration of relatively less tidal heating due to circularization.
This is a subset of case 4 from Section \ref{gen}.
\item
For systems where there is a measured non-zero eccentricity, we simulate the effects of an
eccentricity source by performing the full tidal evolution with an eccentricity floor
equal to the observed value.  These cases are shown in red.
This is essentially a combination of Case 4 and Case 2.
\end{enumerate}
For some planets, some of these ``cases'' were either not possible to compute or in no instances
were the observed parameters consistent with the model parameters.  For instance in cases when the observed
eccentricity is larger than 0.4, the tidal evolution histories with 0.4 maximum initial eccentricity never
are consistent with the observation.  In these cases, no radius range is drawn.
In some of the cases where tidal heating is included, an evolution history is found where
a large amount of energy is deposited into the planet while the orbital parameters are consistent
with observations.  These result in a maximum achieved radius that sometimes exceeds 2 $R_J$.  
In some of these cases, the planet will later cool off before the evolution stops.  
In other cases, the tidal power is sufficient to increase the planet's entropy beyond the
maximum entropy of our grid, which ends the evolutionary calculation. 
In the future we plan to include mass loss and the subsequent evolution history.

By comparing these models we find a few interesting patterns.
When comparing the full tidal evolution model (purple) to the stationary model (blue),
notice that there are some cases where the full tidal model has a larger maximum
radius and other cases where the reverse is true.  This can be understood to be caused
by the two competing effects of tidal evolution.  
Tidal heating puts power into the planet and inflates the radius, 
and tidal orbital evolution allows the planet to cool more efficiently at earlier times when the planet is
less irradiated by the parent star.  
It is also useful to compare these two cases to the no heating model.  The no heating model
generally has a smaller maximum radius than the stationary model because of the second effect.
The tidal model has a larger maximum radius than the no heating model because of energy deposition into 
the planet.  

% Discuss the effects of large initial eccentricity
Often the model achieves large radius values through a recent circularization of an originally high
eccentricity orbit.  During the circularization event  (when the eccentricity drops significantly),
tidal dissipation in the interior of the planet
may deposit sufficient energy to significantly inflate the planet.
The orange case (maximum initial eccentricity equal to 0.4) has been plotted
to compare against the purple (initial eccentricity up to 0.8) to show how large initial eccentricity
evolution histories contribute to the maximum achieved radius.  
Note that in the low \qp\ case in Figure \ref{plrad5}, extremely large radii can be achieved 
for GJ 436b and HAT-P-11.  This happens in our model through a recent rapid circularization of the orbit.  
%These large initial eccentricities may be achieved through planet-planet scattering processes \cp[e.g.][]{ChatterjeeFordMatsumuraRasio2008}.

% Discuss the eccentricity source situations
It may also be possible to have tidal heating without large initial eccentricities
if there is a eccentricity driving source in the system.  In some cases, such as in WASP-6b or WASP-12b, 
the resulting tidal heating may be enough to explain the large transit radius.
By comparing the red (tidal evolution with an eccentricity floor) to the purple (regular tidal evolution), 
larger radius values can be achieved when the orbit is not allowed to circularize.

% conclusion about tidal heating
Tidal evolution and heating clearly have important effects on a planet's evolution, but not all of the
large-radius planets could be explained through this mechanism, given our chosen $Q$ values.
The planets HD 209458b, COROT-EXO-2b, HAT-P-9b, WASP-1b and TrES-4b have radii that are larger
than achieved in our models in both the low and high \qp\ cases.
Typically, while it is possible to inflate the radius to the observed
values, it difficult to find the system with an inflated radius and
low current eccentricity.  WASP-12b was explained if we assume that its eccentricity is maintained.
%This may be possible through interactions with a third body, such as those
%suggested by \citep{Mardling2007}.  

%Comparing figures 5 and 6: begin
When comparing Figure \ref{plrad5} to Figure \ref{plrad65}, 
it is interesting that some of the planets that are not explainable in the lower \qp\ case
can be explained with larger \qp\ . 
%it is apparent 
%that in the second case, many of the planets can have a larger radius range under the full tidal model with the
%larger $Q_{\mathrm{p}}'$ value.  
Although $Q_{\mathrm{p}}'=10^5$ results in tidal heating being stronger than
the $Q_{\mathrm{p}}' = 10^{6.5}$ case, it also results in circularization on a shorter time scale.
In the $Q_{\mathrm{p}}' = 10^{6.5}$ cases, it is often common for there to be a possible recent circularization
of a high initial eccentricity orbit where no such history was found in the $Q_{\mathrm{p}}' = 10^5$ evolution runs.
%end

In Table 1, we have selected a set of the largest planets and listed
various properties.  In the left column, we list the observed parameters.  For
various core sizes, we list the achieved radius of the tidal model in the low \qp\ and high \qp\ cases, 
the estimated luminosity of the planet at its current radius, and
the current contraction rate of the planet without internal heating (previously defined as $\dot{R_{NH}}$).
Also, on the top row for each planet, we list the coefficient of tidal heating.  This is defined as
\begin{eqnarray}
C_T &\equiv& \frac{P_T}{\left(\frac{e}{0.01}\right)^2 \left(\frac{10^5}{Q_{\mathrm{p}}} \right)} \\
    &=& \frac{63}{4} (G M_*)^{3/2} M_* R_p^5 a^{-15/2} \times 10^{-9}
\end{eqnarray}
This quantity allows one to get an order-of-magnitude idea of recent tidal heating 
given the more constrained properties of the system (radius of the planet, masses of the bodies,
and semi-major axis).  The actual tidal power will greatly depend on the eccentricity and $Q$ values, which are more
uncertain.  The ratio between the luminosity of the planet and this coefficient of tidal heating
is a dimensionless number that describes how important tidal effects can be for a given core size.
Certainly, since $P_T \propto e^2$ and \qp\ is quite uncertain, this ratio is not a strong test of tidal effects,
but it is a simple way of testing how important tidal effects presently can be.
Notice also that for an assumed tidal power, we can compute the present contraction rate
using this table and Equation \ref{rdot}.

When calculating the contraction rate, the planet is 
assumed to be located at the current observed semi-major axis, which
determines the incident flux from the star, structure of the planet's atmosphere, and thus the intrinsic luminosity of the planet at each time.
For these large-radius systems, the contraction rate is often very fast.  If we assume that tidal heating is the cause of large
radii, but that an eccentricity driving companion is not present, then either the system
is in a transient period or that this thermal evolution model is not correct.  On the other hand,
if we rule out transient explanations, then either a constant heating is present or it is necessary
to invoke another mechanism.

\subsection{High \qs\ cases}
Although \qs\ is generally thought to be closer to $10^5$ based on the observed circularization time in binaries,
it is possible that that tidal dissipation in the stars is less efficient in the planet-star case.  Since tidal evolution
is not fully understood, the high \qs\ case may or may not be physical.  However, an advantage of this case is
that it allows for orbital history solutions with a recent circularization.  In this regime, the planet migrates
inward at a slower rate and thus the circularization would occur at a later time.  Also, after the tidal power is deposited,
the planet is not rapidly migrating into the star as in the low \qs\ cases.  
\citet{IbguiBurrows2009} have suggested that high \qs\ case can better explain the radius of HD 209458b.  

We have explored this parameter regime as shown in Table 2 for five of the systems that we were not able to explain 
in the low \qs\ cases.  We test the cases \qs\ $= 10^6$ and \qs\ $=10^7$ with both \qp\ $ = 10^5$ and \qp\ $ = 10^{6.5}$.  
In the table the radius range is reported
for a given core size, \qp\ and \qs\ model parameters, as well as the number of runs that were found at some point in 
time to be consistent with the observed age, semi-major axis and eccentricity of the system.  

\begin{figure}
  \begin{center}
  \subfloat[] {\label{HD209458gridA} \includegraphics[width=0.9\linewidth]{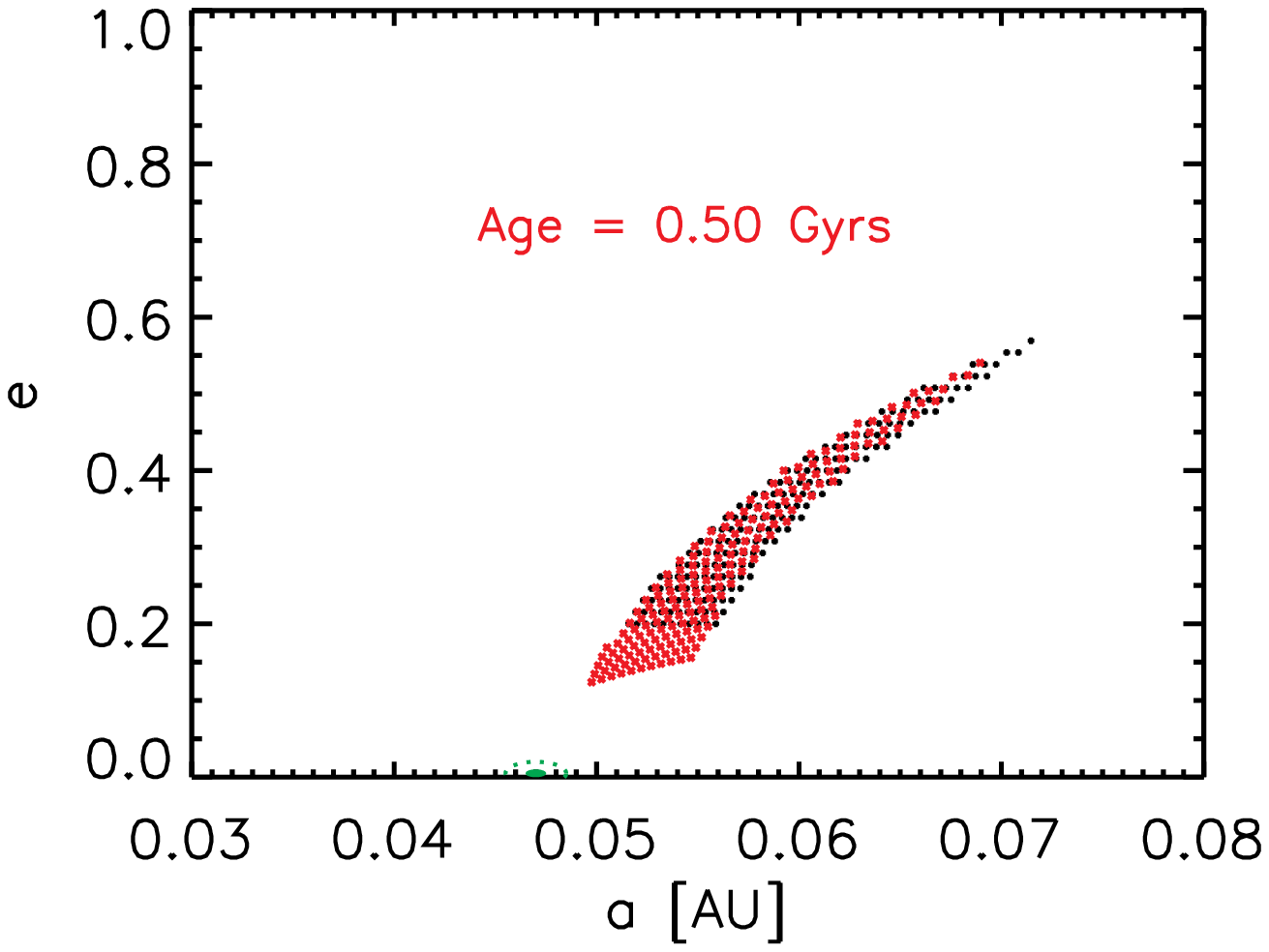}} \\
  \subfloat[] {\label{HD209458gridB} \includegraphics[width=0.9\linewidth]{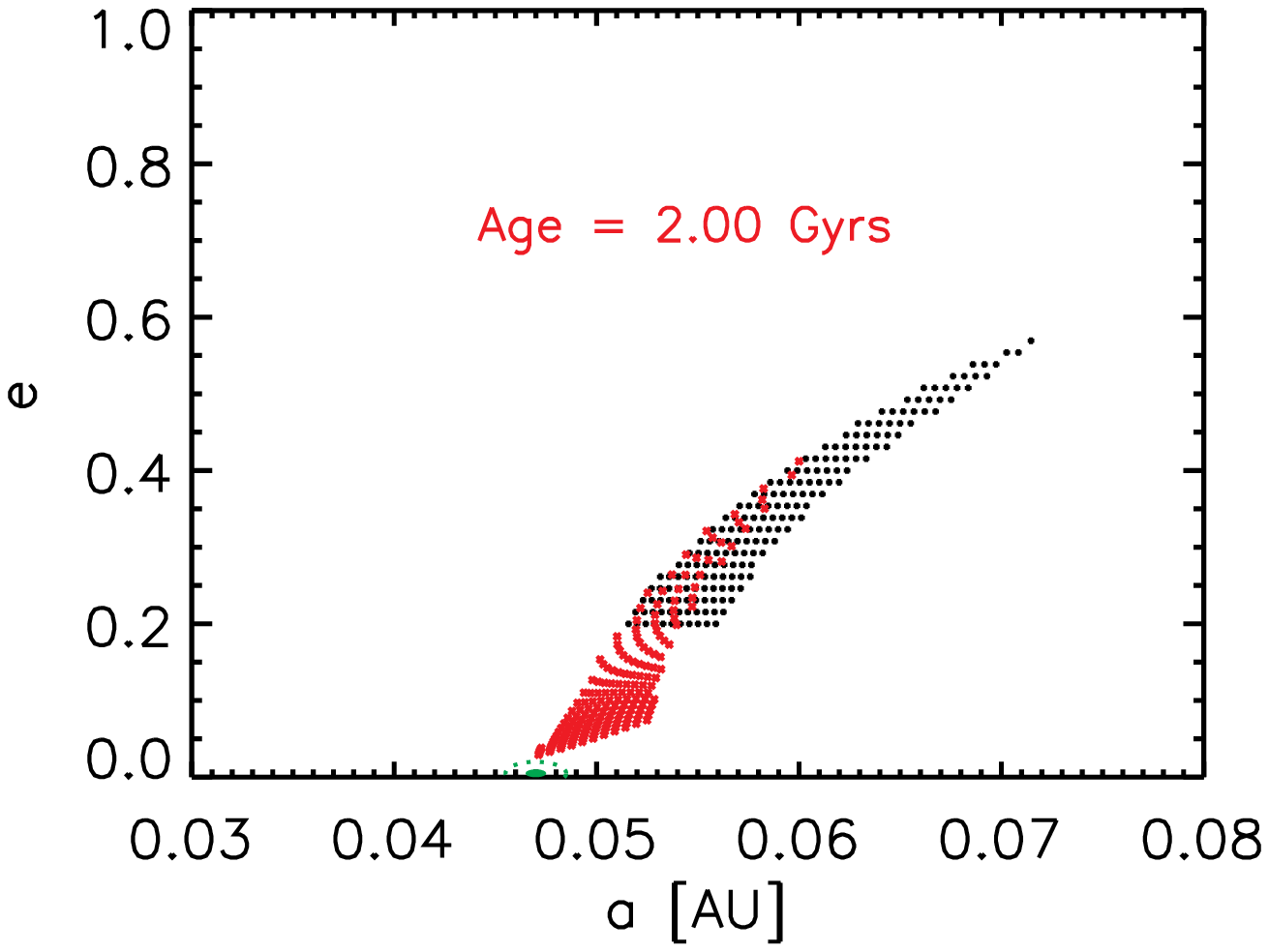}} \\
  \subfloat[] {\label{HD209458gridC} \includegraphics[width=0.9\linewidth]{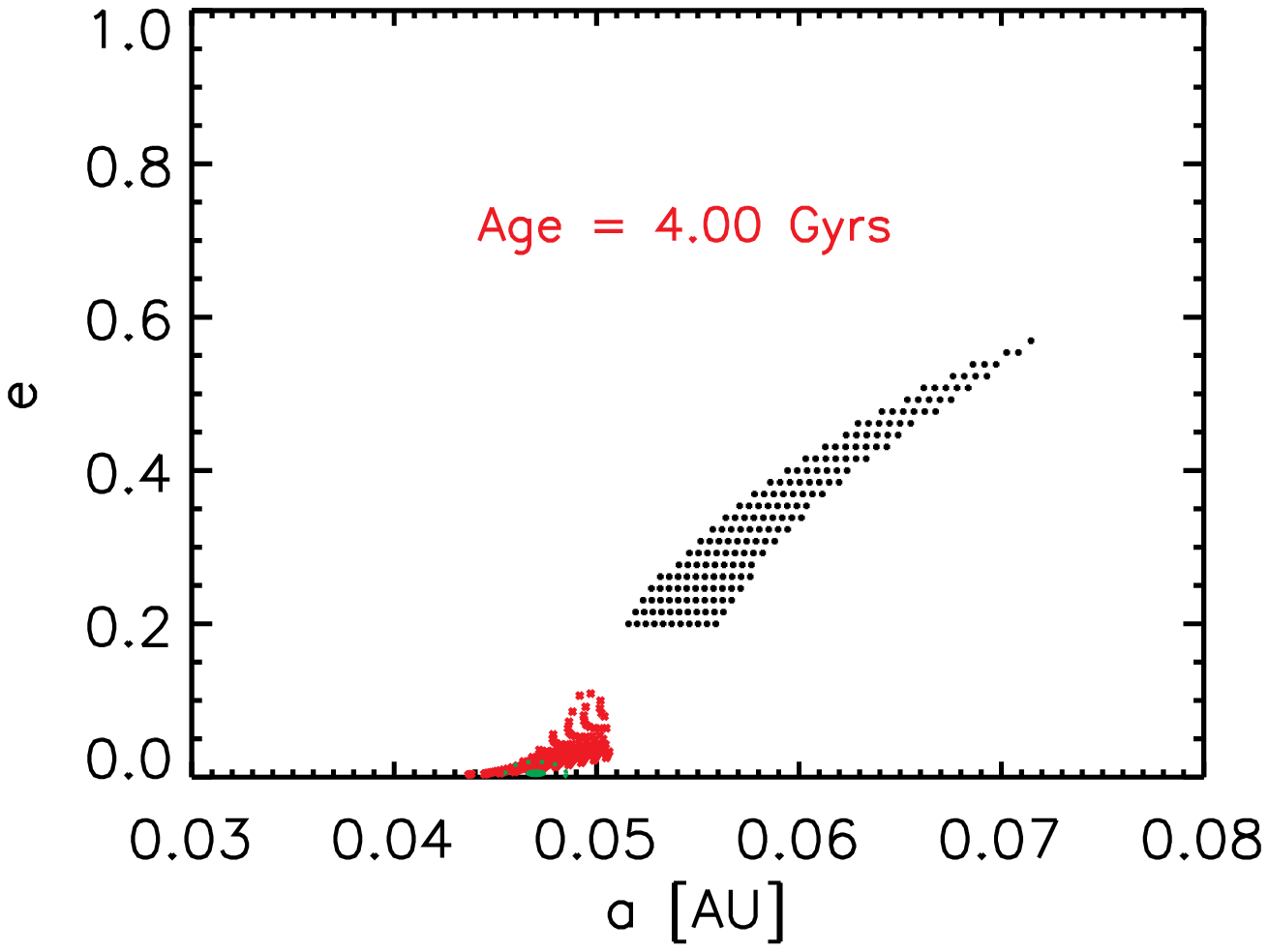}} 
  \caption{    
    Grid of evolution histories (with initial $e > 0.2$) that were found to be consistent with the orbital 
    parameters at a later time for the system HD 209458.
    These evolution runs assume there is no core, \qp\ $=10^{6.5}$ and \qs\ $=10^6$.
    Black: original orbital parameters of each run.  Red: orbital parameters at a later marked time (0.5 Gyr, 1.5 Gyr, and 2.1 Gyr).
    The filled green circle is the 1 $\sigma$ zone, while the dashed region is the 3 $\sigma$ zone.
  }
  \label{HD209458grid}
  \end{center}
\end{figure}

Also, in Figure \ref{HD209458grid}, we show snapshots in semi-major axis / eccentricity space of possible evolution histories
of HD 209458 b 
that are consistent with the observed parameters.  The black points are the original orbital parameters, while
the red points are the orbital parameters at a later time.  The green oval is the 1 $\sigma$ orbital parameters.
The dashed green line is the 3 $\sigma$ orbital parameters, which we require an evolution histories to fall within
during the expected age range of the system.  Eccentricity was sampled from 0.2 to 0.8 in this particular case.

\begin{figure}
\includegraphics[width=0.94\linewidth]{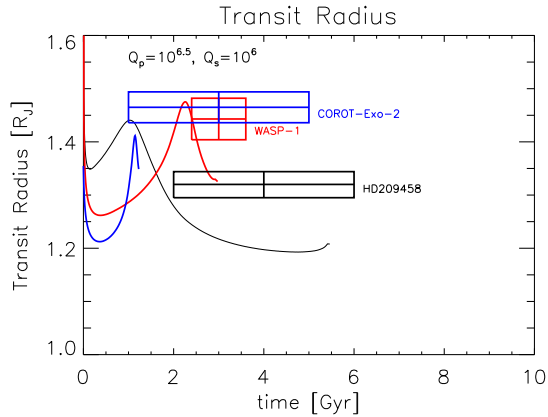} 
  \caption{
    Potential radius evolution histories for HD 209458b, WASP-1b, and CoRoT-Exo-2b with no core, 
    \qp\ $=10^{6.5}$ and \qs\ $=10^6$ (larger than our standard case).
    As usual, these evolution histories have been selected from an ensemble of possible initial conditions such that
    at some point during the estimated age of the system, the planet has orbital parameters that are consistent with the observed values.
}
\label{hqrad}
\end{figure}

We also show in Figure \ref{hqrad} possible radius evolution histories for the planets
HD 209458b, WASP-1b, and CoRoT-Exo-2b.  When \qs\ is allowed to be larger, the qualitative
effect is that the planet's semi-major axis decreases slower and thus the circularization
event occurs at a later time.  This makes it possible to sometimes achieve higher radius 
values at the expected age of the system with the model.  However, even for these high \qs\ runs for these large-radius planets, only for two of the five can the observed radius be matched.

\section{Discussion \& Conclusions}
This paper presents a coupled tidal and thermal evolution model applicable to close-in extrasolar giant planets.
The model is tested against 45 of the known transiting systems.  
Generally, tidal evolution yields two competing effects on the radii of close-in EGPs:
\begin{enumerate}
\item
  Tidal evolution requires that, after planet formation and subsequent fast migration to a relatively close-in orbit, 
  the planet start at a larger semi-major axis than is currently observed \cp{Jackson08a}.
  This results in less incident flux at earlier times, which allows the planet
  to cool more efficiently and contract more at a young age, which
  moves the range of feasible model radii at the current time
  to smaller values.  
  Generally this is a minor effect, but it is more important for cases when the
  current incident flux is larger.  
\item
  Tidal evolution deposits energy into the planet when the orbit is being circularized.
  This typically increases the radius of the planet at this time.  If there is an eccentricity
  driving source for the inner planet, then tidal heating can be important for the duration of the planet's
  life.  If the planet starts with a highly eccentric orbit, it might not circularize for gigayears.  The semi-major axis
  of the planet's orbit will initially slowly decrease due to tides on the star.  As the planet moves closer
  to the star, tides on the planet become more effective.  This delay of circularization can sometimes allow
  tidal heating to significantly inflate planets multiple gigayears after formation despite these systems having
  shorter ``circularization'' time scales.  
% until the semi-major
%  axis of the planet decreases enough such that tides on the planet become effective.
%
%, which may 
%  increase the radius for a short period.  We define the ``circularization'' event as 
%  when the orbit is being transformed from an eccentric orbit to a circular orbit and when
%  when tidal power is being deposited into the planet.
%  This time scale depends on the physical parameters, $M_{\mathrm{p}}$, $M_{\mathrm{s}}$, $a$, the model parameters,
%  \qp\ and \qs\, and the initial orbital parameters $a_0$ and $e_0$.  
%  Note that this tidal ``circularization'' does not necessarily start after the 
%  planet is born with an eccentric orbit.  The tidal effects can be relatively
%  unimportant for a time scale on the order of these system ages before the
%  circularization occurs.  During this period the semi-major axis will slowly
%  decrease until tidal evolution becomes important.  Then tidal effects can accelerate
%  and the circularization occurs.
%  If the time to circularize is short then a tidal heating explanation for inflated planets
%%  would require that these planets are being observed at a very special time.  
%  Also important is how large the luminosity of the planet is relative to the possible tidal
%  power.  If the possible tidal power is always significantly smaller than the
%  luminosity of the planet, then tidal evolution can not inflate the planet 
%  even with an eccentricity driving source.  
\end{enumerate}

We have shown that for the close-in giant planets that orbital history can play a large role in
determining the thermal evolution and current observed radius.  While the effects are larger 
for planets with larger initially eccentricities, tidal evolution still affects the thermal 
evolution of planets with zero eccentricity as well.
Varying amounts of time-dependent tidal heating are degenerate with the radius effects due 
to the core of a planet (or more generally, a heavy element enrichment).

Since at the current time we are ignorant of the exact orbital history, it is generally not 
possible to determine the mass of the core with complete confidence for any specific system.  
However, in cases when the radius of the planet is especially small, a large core or increased heavy element 
abundance is required.  For larger radius planets, it is not possible to determine the planet's core size because
recent tidal heating is degenerate with smaller core sizes.
Furthermore, some systems likely have more complex orbital dynamics than described here due 
to the effects a third body.  The uncertainty is increased since despite our expectation that tidal effects do occur, the
rate that at which they occur (controlled by $Q$) is uncertain to an order of magnitude.

This paper serves as a \emph{forward} test of the tidal theory for close-in EGPs outlined by \citet{Jackson08}, 
who had previously only investigated heating rates \emph{backwards} in time, from current small eccentricities from 0.001 to 0.03.  
Quite often however, the forward modeling of these single-planet systems, across a wide swath of initial $a$ and $e$, is not 
consistent with current eccentricities as large as \ct{Jackson08} assumed.  
If initial eccentricities were indeed large, then final circularization and tidal surge may indeed by fairly recent, 
but this cannot be expected to be the rule in these systems.  
We have taken an agnostic view as to whether initial migration to within 0.1 AU was via scattering or disk migration.  
In the former, initial eccentricities up to 0.8 are possible \cp{ChatterjeeFordMatsumuraRasio2008} 
while in the latter the initial eccentricity would be zero.  The viability of tidal heating to explain 
even some of the inflated planets with very small \emph{current} eccentricities rests on the notion that 
planet scattering does occur, such that circularization (and radius inflation) can occur at gigayear ages.  
The detection of misalignment between the planetary orbital plane axis and stellar rotation axis via the 
Rossiter-McLaughlin Effect \cp[e.g.][]{Winn07c,Winn08} is beginning to shed light on migration.  
\ct{Fabrycky09} have found tentative evidence that is consistent with two modes of migration, 
one which may yield close alignment (perhaps from disk migration) and one with which may yield random alignment 
(perhaps from scattering), although to date only XO-3b in the published literature shows a large misalignment \cp{Hebrard08}.  
Further measurements will help to constrain the relative importance of these two modes of migration.

Most of the systems investigated do not require tidal heating to match their radius, 
but these systems can also be readily explained
when including tidal evolution.  Some of the planets investigated can be matched with tidal heating that 
could not be explained with a standard contraction model.  Depending on the \qp\ value chosen, 
HAT-P-4, HAT-P-9, XO-4, HAT-P-6, OGLE-TR-211, WASP-4, WASP-12, TrES-3, HAT-P-7, and OGLE-TR-56
can all be explained with an evolution history with non-zero initial eccentricity.  WASP-6 and WASP-12
can be explained by invoking a minimum eccentricity, which may suggest the presence of a companion.
Other systems were not explained by the model for our chosen \qp\ values.  This suggests that either
\qp\ and \qs\ may be much different then our expectation or that other mechanisms are at work 
in these large-radius planets.

% Modify this paragraph
This work should be taken as a simplified analysis of how tidal evolution can affect a planet's thermal evolution.
Strong quantitative conclusions should not be drawn because of the large uncertainties in the tidal evolution model, especially
at large eccentricity.
Also, the rate of tidal effects may be a very strong function of frequency.  If this is the case, the planet may spend 
a lot of time at certain states where tidal effects are slow and rapidly pass through states where tidal effects are more rapid.  
If a constant Q value can even be applied, the actual value is highly uncertain.  The Q values that we choose were meant only
to span the range that we considered to be likely.  
The rate of tidal effects may depend on the interior
structure of the planet and may be different for different exoplanets.  
Also, this analysis only takes into account orbit-circularization tidal heating.

The conclusion that should be drawn from this work is that a planet's tidal evolution history can play an important role 
on the planets' current radius, especially for systems that are born at semi-major axis less than 0.1 AU.  
In some cases, tidal heating could have inflated the radius of the planet in the recent past, even 
though tidal heating in the present might not be happening.  In other cases, we were not able to explain the 
large-radius observations
with our coupled tidal-thermal evolution model.  This suggests that tidal heating will not be able to explain all of the 
large-radius planets, which has been a hope of some authors \cp{Jackson08,IbguiBurrows2009}.
For some of the planets that we are able to explain, we require a recent circularization, 
such that this model can only explain these observations if we at at a ``special time'' in its evolution. 
This has to be reconciled with the fraction of planets that have large radii that require such an explanation.
Improved constraints on the eccentricities of these systems will better constrain recent tidal heating.

A more robust treatment of the effects of tidal heating on transiting planet radius evolution may require a 
coupling of the model presented here to a scattering/disk migration model, which could derive the statistical 
likelihood of various initial orbital $a$ and $e$ configurations, which would then serve as the initial conditions 
to subsequent orbital-tidal and thermal evolution.  This is important because for any particular planetary system 
the orbital evolutionary history of the close-in planet may be difficult to ascertain.  
Recently \ct{Nagasawa08} have simulated the formation of hot Jupiters with a coupled scattering and tidal evolution code, 
and find a frequent occurence of hot Jupiter planets.  A further coupled undertaking of this sort, to be compared with an 
statistically significant number of transiting planets, could be performed in the future.

JJF and NM are supported by NSF grant AST-0832769.  We thank the referee, G. Chabrier, as well as E. Ford, D. Fabrycky, 
and S. Gaudi for their comments.

\bibliography{references}
\clearpage
%----------------------------------------------------------------------------------------------

\begin{center}
\begin{deluxetable}{lcccrc}
\tablecaption{MODEL CALCULATIONS FOR SELECTED TRANSITING SYSTEMS}
\tablehead{
  \colhead{\textbf{System}} &
  \colhead{\textbf{Core} [$M_E$]} &
  \colhead{\textbf{Radius Range (\qp\ $=10^5$)}} &
  \colhead{\textbf{Radius Range (\qp\ $=10^{6.5}$)}} &
  \colhead{\textbf{$P$ [ergs/s]}} &
  \colhead{\textbf{$\dot{R}_{NH}$ [$R_J$/yr]}}
}
\tablecolumns{6}
%\tablecaption{
%Various large-radius hot Jupiter planets have been listed.
%On the left, we list the observed parameters of the system for reference.
%In the first row for each planet in the $P_{\textrm{eq}}$ column, we list the coefficient of tidal power.
%The achieved radius range is listed for the two different \qp\ values.
% $P_{eq}$ is the required equilibrium internal heating that would be required to maintain the radius in steady state.
%$\dot{R}$ is the contraction rate without internal heating.
%}
\label{PlanetTable}
\tablewidth{450pt}
\startdata
\textbf{HD209458} & & & &  $ C_T = 6.3\times 10^{25}$ & \\
\hline
$M_p$ =  0.69 $M_J$ &    0.0 &  1.12 -  1.19 &  1.13 -  1.18 & $ L= 1.5\times 10^{26}$ &$   -4.2\times 10^{-7}$\\
$R_p$ = 1.32 $R_J$ &   10.0 &  1.08 -  1.15 &  1.08 -  1.15 & $ L= 3.8\times 10^{26}$ &$  - 1.\times 10^{-6}$\\
a = 0.05 AU &  30.0 &  1.02 -  1.08 &  1.02 -  1.07 & $ L= 1.6\times 10^{27}$ &$  - 4.5\times 10^{-6}$\\
$e = 0.00$ &  100.0 &  0.81 -  0.90 &  0.81 -  0.84 & $ L= 7.6\times 10^{28}$ &$ -  1.5\times 10^{-4}$\\
\hline
\textbf{COROT-Exo-1} & & & & $ C_T = 9.2\times 10^{27}$ & \\
\hline
$M_p$ =  1.03 $M_J$ &    0.0 &  1.14 -  1.23 &  1.16 -  1.79 & $ L= 1.2\times 10^{27}$ &$  - 2.2\times 10^{-6}$\\
$R_p$ = 1.49 $R_J$ &   10.0 &  1.11 -  1.21 &  1.13 -  1.79 & $L=  1.8\times 10^{27}$ &$  - 3.3\times 10^{-6}$\\
a = 0.03 AU &  30.0 &  1.07 -  1.15 &  1.08 -  1.52 & $ L= 3.8\times 10^{27}$ &$  - 7.3\times 10^{-6}$\\
$e = 0.00$ &  100.0 &  0.95 -  1.03 &  0.93 -  1.07 & $ L= 5.3\times 10^{28}$ &$  - 1.\times 10^{-4}$\\
\hline
\textbf{COROT-Exo-2} & & & & $ C_T = 3.8\times 10^{27}$ & \\
\hline
$M_p$ =  3.31 $M_J$ &    0.0 &  1.11 -  1.23 &  1.11 -  1.17 & $ L= 6.1\times 10^{28}$ &$  - 1.4\times 10^{-5}$\\
$R_p$ = 1.47 $R_J$ &   10.0 &  1.11 -  1.24 &  1.11 -  1.16 & $ L= 7.0\times 10^{28}$ &$  - 1.6\times 10^{-5}$\\
a = 0.03 AU &  30.0 &  1.09 -  1.23 &  1.09 -  1.15 & $ L= 8.7\times 10^{28}$ &$  - 2.\times 10^{-5}$\\
$e = 0.00$ &  100.0 &  1.05 -  1.20 &  1.05 -  1.10 & $ L= 1.6\times 10^{29}$ &$  - 3.9\times 10^{-5}$\\
\hline
\textbf{XO-4} & & & & $ C_T = 3.3\times 10^{25}$ & \\
\hline
$M_p$ =  1.72 $M_J$ &    0.0 &  1.15 -  1.34 &  1.15 -  1.17 & $  L= 1.7\times 10^{27}$ &$  - 8.9\times 10^{-7}$\\
$R_p$ = 1.34 $R_J$ &   10.0 &  1.14 -  1.30 &  1.13 -  1.15 & $  L= 2.4\times 10^{27}$ &$  - 1.2\times 10^{-6}$\\
a = 0.06 AU &  30.0 &  1.11 -  1.25 &  1.10 -  1.13 & $  L= 4.2\times 10^{27}$ &$  - 2.2\times 10^{-6}$\\
$e = 0.00$ &  100.0 &  1.02 -  1.11 &  1.02 -  1.03 & $  L= 3.2\times 10^{28}$ &$  - 1.9\times 10^{-5}$\\
\hline
\textbf{HAT-P-6} & & & & $C_T =  4.3\times 10^{25}$ & \\
\hline
$M_p$ =  1.06 $M_J$ &    0.0 &  1.16 -  1.29 &  1.16 -  1.19 & $ L= 4.2\times 10^{26}$ &$  - 4.7\times 10^{-7}$\\
$R_p$ = 1.33 $R_J$ &   10.0 &  1.14 -  1.28 &  1.13 -  1.16 & $ L= 7.2\times 10^{26}$ &$  - 8.3\times 10^{-7}$\\
a = 0.05 AU &  30.0 &  1.09 -  1.28 &  1.09 -  1.11 & $ L= 1.7\times 10^{27}$ &$  - 2.1\times 10^{-6}$\\
$e = 0.00$ &  100.0 &  0.95 -  1.09 &  0.95 -  0.96 & $ L= 3.2\times 10^{28}$ &$  - 4.7\times 10^{-5}$\\
\hline
\textbf{HAT-P-7} & & & & $ C_T = 8.0\times 10^{26}$ & \\
\hline
$M_p$ =  1.78 $M_J$ &    0.0 &  1.14 -  1.55 &  1.14 -  1.21 & $ L= 6.3\times 10^{26}$ &$  - 3.2\times 10^{-7}$\\
$R_p$ = 1.36 $R_J$ &   10.0 &  1.13 -  1.56 &  1.12 -  1.19 & $ L= 8.3\times 10^{26}$ &$  - 4.3\times 10^{-7}$\\
a = 0.04 AU &  30.0 &  1.11 -  1.50 &  1.10 -  1.16 & $ L= 1.4\times 10^{27}$ &$ -  7.3\times 10^{-7}$\\
$e = 0.00$ &  100.0 &  1.01 -  1.44 &  1.02 -  1.06 & $ L= 6.8\times 10^{27}$ &$ -  4.2\times 10^{-6}$\\
\hline
\textbf{HAT-P-9} & & & & $ C_T = 5.0\times 10^{25}$ & \\
\hline
$M_p$ =  0.78 $M_J$ &    0.0 &  1.16 -  1.49 &  1.16 -  1.29 & $ L= 7.0\times 10^{26}$ &$ -  1.7\times 10^{-6}$\\
$R_p$ = 1.40 $R_J$ &   10.0 &  1.13 -  1.50 &  1.13 -  1.25 & $ L= 1.3\times 10^{27}$ &$  - 3.3\times 10^{-6}$\\
a = 0.05 AU &  30.0 &  1.06 -  1.36 &  1.06 -  1.17 & $ L= 3.7\times 10^{27}$ &$ -  1.\times 10^{-5}$\\
$e = 0.00$ &  100.0 &  0.87 -  1.00 &  0.87 -  0.95 & $ L= 8.6\times 10^{28}$ &$ -  1.7\times 10^{-4}$\\
\hline
\textbf{TrES-4} & & & & $ C_T = 3.9\times 10^{26}$ & \\
\hline
$M_p$ =  0.93 $M_J$ &    0.0 &  1.15 -  1.33 &  1.14 -  1.17 & $ L= 1.0\times 10^{28}$ &$  - 4.4\times 10^{-5}$\\
$R_p$ = 1.78 $R_J$ &   10.0 &  1.12 -  1.32 &  1.11 -  1.14 & $ L= 1.4\times 10^{28}$ &$  - 6.\times 10^{-5}$\\
a = 0.05 AU &  30.0 &  1.07 -  1.29 &  1.06 -  1.09 & $ L= 3.4\times 10^{28}$ &$  - 1.2\times 10^{-4}$\\
$e = 0.00$ &  100.0 &  0.91 -  0.99 &  0.90 -  0.92 & - & - \\
\hline
\textbf{OGLE-TR-211} & & & & $ C_T = 6.4\times 10^{25}$ & \\
\hline
$M_p$ =  1.03 $M_J$ &    0.0 &  1.14 -  1.38 &  1.14 -  1.22 & $ L= 5.0\times 10^{26}$ &$  - 6.7\times 10^{-7}$\\
$R_p$ = 1.36 $R_J$ &   10.0 &  1.12 -  1.36 &  1.12 -  1.19 & $ L= 8.2\times 10^{26}$ &$   -1.1\times 10^{-6}$\\
a = 0.05 AU &  30.0 &  1.08 -  1.38 &  1.07 -  1.13 & $ L= 1.9\times 10^{27}$ &$  - 2.7\times 10^{-6}$\\
$e = 0.00$ &  100.0 &  0.93 -  1.10 &  0.93 -  0.97 & $ L= 3.6\times 10^{28}$ &$ -  5.8\times 10^{-5}$\\
\hline
\textbf{WASP-1} & & & & $ C_T = 5.2\times 10^{26}$ & \\
\hline
$M_p$ =  0.87 $M_J$ &    0.0 &  1.16 -  1.25 &  1.16 -  1.21 & $  L= 6.1\times 10^{26}$ &$  - 1.4\times 10^{-6}$\\
$R_p$ = 1.44 $R_J$ &   10.0 &  1.13 -  1.22 &  1.13 -  1.18 & $  L= 1.0\times 10^{27}$ &$  - 2.4\times 10^{-6}$\\
a = 0.04 AU &  30.0 &  1.07 -  1.18 &  1.07 -  1.10 & $  L= 2.4\times 10^{27}$ &$  - 6.2\times 10^{-6}$\\
$e = 0.00$ &  100.0 &  0.90 -  1.06 &  0.90 -  0.92 & $  L= 5.1\times 10^{28}$ &$ -  1.2\times 10^{-4}$\\
\hline
\textbf{WASP-4} & & & & $ C_T = 1.3\times 10^{28}$ & \\
\hline
$M_p$ =  1.27 $M_J$ &    0.0 &  1.12 -  1.20 &  1.13 -  1.66 & $ L= 3.3\times 10^{27}$ &$  - 3.8\times 10^{-6}$\\
$R_p$ = 1.45 $R_J$ &   10.0 &  1.11 -  1.18 &  1.10 -  1.51 & $ L= 4.6\times 10^{27}$ &$  - 5.4\times 10^{-6}$\\
a = 0.02 AU &  30.0 &  1.07 -  1.11 &  1.08 -  1.52 & $ L= 8.2\times 10^{27}$ &$  - 1.\times 10^{-5}$\\
$e = 0.00$ &  100.0 &  0.96 -  1.03 &  0.96 -  1.18 & $ L= 7.5\times 10^{28}$ &$  - 8.7\times 10^{-5}$\\
\hline
\textbf{WASP-12} & & & & $ C_T = 1.1\times 10^{29}$ & \\
\hline
$M_p$ =  1.41 $M_J$ &    0.0 &  - &  1.18 -  2.02 & $ L= 2.5\times 10^{28}$ &$ -  5.2\times 10^{-5}$\\
$R_p$ = 1.79 $R_J$ &   10.0 &  - &  1.16 -  1.57 & $ L= 3.6\times 10^{28}$ &$  - 7.\times 10^{-5}$\\
$a = 0.02$ AU &  30.0 &  - &  1.12 -  1.37 & $ L= 5.9\times 10^{28}$ &$  - 1.1\times 10^{-4}$\\
$e = 0.05$ &  100.0 &  - &  1.01 -  1.11 & - & - \\
\hline
\enddata
\tablecomments{Various large-radius hot Jupiter planets have been listed.
In the first column, we list the observed parameters of the system for reference.
In the second column, we list an assumed core size.  The achieved radius range for two different
\qp\ values is liested in the third and fourth columns.
In the fifth column, we list relevant power quantities.  The coefficient of tidal 
power is listed in the first row for each system.  In the following rows, we list the luminosity
of the planet for the assumed core mass.  In the final row, we calculate $\dot{R}_{NH}$, the radius
derivative when there is no internal heating source.
}
\end{deluxetable}
\end{center}

\clearpage
\begin{center}
\begin{deluxetable}{lccccc}
\tablehead{
\colhead{\textbf{System}} & 
\colhead{\textbf{Core} [$M_E$]} &
\colhead{\textbf{Radius} [$R_J$] (5,6)} &
\colhead{\textbf{Radius} [$R_J$] (5,7)} &
\colhead{\textbf{Radius} [$R_J$] (6.5,6)} &
\colhead{\textbf{Radius} [$R_J$] (6.5,7)}
}
\tablewidth{450pt}
\startdata
\textbf{HD209458} & & & &  & \\
\hline
$M_p$ =  0.69 $M_J$ &    0.0 &  1.12 -  1.19 ( 683 )  &  1.12 -  1.18 ( 737 )  &  1.15 -  1.32 ( 816 )  &  1.15 -  1.31 (1036 )  \\
$R_p$ = 1.32 $R_J$ &   10.0 &  1.09 -  1.16 ( 931 )  &  1.09 -  1.15 (1136 )  &  1.12 -  1.27 ( 765 )  &  1.11 -  1.25 ( 945 )  \\
\hline
\textbf{TrES-4} & & & &  & \\
\hline
$M_p$ =  0.93 $M_J$ &    0.0 &  1.16 -  1.22 (1291 )  &  1.16 -  1.21 ( 849 )  &  1.24 -  1.43 ( 665 )  &  1.19 -  1.37 (1205 )  \\
$R_p$ = 1.78 $R_J$ &   10.0 &  1.13 -  1.19 (1285 )  &  1.13 -  1.18 ( 959 )  &  1.20 -  1.37 ( 512 )  &  1.16 -  1.33 (1154 )  \\
\hline
\textbf{HAT-P-8} & & & &  & \\
\hline
$M_p$ =  1.52 $M_J$ &    0.0 &  1.15 -  1.19 (1520 )  &  1.15 -  1.19 (1390 )  &  1.17 -  1.28 ( 538 )  &  1.18 -  1.30 ( 728 )  \\
$R_p$ = 1.58 $R_J$ &   10.0 &  1.13 -  1.18 (1515 )  &  1.13 -  1.18 (1390 )  &  1.16 -  1.26 ( 501 )  &  1.17 -  1.28 ( 694 )  \\
\hline
\textbf{WASP-1} & & & &  & \\
\hline
$M_p$ =  0.87 $M_J$ &    0.0 &  1.17 -  1.21 ( 835 )  &  1.17 -  1.20 (  26 )  &  1.23 -  1.48 ( 656 )  &  1.19 -  1.39 (1463 )  \\
$R_p$ = 1.44 $R_J$ &   10.0 &  1.14 -  1.18 ( 829 )  &  1.14 -  1.17 ( 297 )  &  1.20 -  1.45 ( 636 )  &  1.16 -  1.35 (1438 )  \\
\hline
\textbf{COROT-Exo-2} & & & &  & \\
\hline
$M_p$ =  3.31 $M_J$ &    0.0 &  1.12 -  1.18 (1337 )  &  1.12 -  1.19 (1069 )  &  1.19 -  1.40 (1243 )  &  1.13 -  1.33 (2127 )  \\
$R_p$ = 1.47 $R_J$ &   10.0 &  1.11 -  1.17 (1334 )  &  1.11 -  1.19 (1092 )  &  1.18 -  1.39 (1242 )  &  1.12 -  1.32 (2120 )  \\
\hline
\enddata
\tablecomments{Achieved radius values for 5 systems with high \qs\ for core size 0.0 and 10 \me\ .  
  The parameters used are denoted in the header with ($\log$ \qp\, $\log$ \qs\ ).  In the body of the
  table, the range or achieved radius values is lested along with the number of runs found in parenthesis.  
}
\end{deluxetable}
\end{center}

\end{document}